\documentclass[a4paper,11pt]{article}
\usepackage{jheppub} 
\usepackage{array}
\usepackage{accents}
\newcommand{\rsq}[1]{|#1]}

\newcommand{\lan}[1]{\langle#1|}
\newcommand{\sq}[1]{[#1]}
\newcommand{\an}[1]{\langle#1\rangle}
\newcommand{\mo}{\mathcal{O}}
\newcommand{\mA}{\mathcal{A}}
\newcommand{\mAt}{\tilde{\mathrm{A}}}
\newcommand{\zb}{\bar{z}}
\newcommand{\hypf}{{}_2F_{1}}
\newcommand{\D}{\Delta}
\newcommand{\e}{\epsilon}
\newcommand{\om}{\omega}
\newcommand{\tom}{\tilde{\omega}}
\newcommand{\shat}{\hat{s}}
\newcolumntype{L}{>{$}l<{$}}

\title{Scalar-Graviton Amplitudes and Celestial Holography}
\author[a,b]{Adam Ball,}
\author[a]{Shounak De,}
\author[a,c]{Akshay Yelleshpur Srikant}
\author[a,d]{and Anastasia Volovich}

\affiliation[a]{Department of Physics,
	Brown University,
	Providence, RI 02912, USA}
 \affiliation[b]{Perimeter Institute for Theoretical Physics, Waterloo, ON N2L 2Y5, Canada}
 \affiliation[c]{Mathematical Institute, University of Oxford, Oxford, OX2 6GG, UK}
\affiliation[d]{Department of Physics,
Harvard University,
Cambridge, MA 02138, USA }

\emailAdd{aball1@perimeterinstitute.ca}
\emailAdd{shounak\_de@brown.edu}
\emailAdd{Akshay.YelleshpurSrikant@maths.ox.ac.uk}
\emailAdd{anastasia\_volovich@brown.edu}

\abstract{We compute scattering amplitudes involving one massive scalar and two, three, or four gravitons. We show that when the conformal dimension of the massive scalar is set to zero, the resulting celestial correlators depend {\it only} on the coordinates of the gravitons. Such correlators of gravitons are well-defined and do not suffer from divergences associated with the Mellin transform of usual graviton amplitudes. Moreover, they are non-distributional and take the form of standard CFT correlators. We show that they are consistent with the usual OPEs but the statement of the soft theorem is modified.}

\begin{document}
\maketitle
	
\section{Introduction}
\label{sec:Intro}
Understanding how bulk translational symmetry emerges is one of the central problems
in the celestial holography program. In addition to being constrained by conformal invariance, 
correlation functions in celestial conformal field theory (CCFT) \cite{Pasterski:2016qvg, Pasterski:2017kqt, Pasterski:2017ylz} must also obey translation invariance. This acts as an ``external symmetry" and forces low point correlators to be distributional \cite{Law:2019glh}. These effects are not as stark at higher points where the correlators are non-distributional. However, the positions of the operators are still constrained and cannot be chosen to lie arbitrarily on the celestial sphere \cite{Mizera:2022sln}. Such correlation functions are exotic from the perspective of a 2D conformal field theory and are one of the main roadblocks to finding a top-down description of CCFT. Indeed many of the existing top-down constructions \cite{Costello:2022jpg, Costello:2023hmi, Bittleston:2023bzp} correspond to theories in the bulk which differ from Yang-Mills or GR in significant ways. \\

One way to circumvent this difficulty has been to introduce nontrivial backgrounds in the bulk \cite{Casali:2022fro, Fan:2022kpp, Fan:2022vbz, Stieberger:2022zyk, Gonzo:2022tjm, Banerjee:2023rni, Adamo:2020syc, Adamo:2020yzi, Adamo:2022mev, Adamo:2023fbj, Adamo:2023zeh, Bogna:2023bbd}. The backgrounds must not only be Lorentz invariant --- thus preserving the 2D conformal symmetry of the CCFT --- but should also maintain the asymptotic flatness of the 4D bulk spacetime. The celestial counterparts of scattering amplitudes on these backgrounds typically have a number of desirable features. Firstly, they are unconstrained by translation invariance: the background acts as a source/sink of momentum. Consequently, they resemble traditional CFT correlators. Secondly, they have better convergence properties and we avoid having to deal with divergent integrals. Finally, for well-chosen backgrounds, the OPEs and symmetry algebras of the resulting theory resemble those of the CCFT without a background \cite{Casali:2022fro, Banerjee:2023rni, Melton:2022fsf}. These properties suggest that the CFT from which these correlators arise might be closely related to the background free CCFT. Moreover, the semblance of these correlators to those of a traditional CFT makes it easier to identify an independent definition. We refer the reader to \cite{Stieberger:2022zyk, Stieberger:2023fju} for some progress in this direction in the case of Yang-Mills theory. \\

In this work, we are concerned with gravity amplitudes in 4D Minkowski and Klein spacetimes \cite{Atanasov:2021oyu}. In particular, our interests lie in generalizing the results of \cite{Melton:2022fsf}, where the authors considered gluon amplitudes on a massive scalar background, to gravity amplitudes. The scalar was not colored and interacted with gluons only via a non-minimal coupling which gave rise to particularly simple amplitudes. A straightforward extension of this to gravity is impossible since a non-dynamical scalar background breaks diffeomorphism invariance. Moreover, it is impossible to avoid a minimal coupling of the scalar to gravitons. Thus, amplitudes involving $n$ gravitons and one scalar will necessarily involve a scalar exchange. Nevertheless, we will show that the celestial counterparts of these amplitudes exhibit all of the desirable properties mentioned in the previous paragraph. \\

This paper is structured as follows. In Section \ref{sec:perlims}, we describe the precise quantity of interest and its relation to scattering amplitudes. In Section \ref{sec:scamps}, we compute scattering amplitudes involving two, three, and four gravitons and a single massive scalar. To our knowledge, these amplitudes have not appeared in the literature before. We then use these to compute the associated celestial correlators in Section \ref{sec:celestialscalargravamp}. The OPE of two positive helicity gravitons extracted from these correlators is consistent with the usual OPE as shown in Section \ref{sec:OPE}. However, the celestial avatar of the leading soft theorem appears to be slightly different, which we expand on in Section \ref{sec:conformalsofttheorem}. Finally, we end with some outlook and discussions in Section \ref{sec:disc}.

\section{Preliminaries}
\label{sec:perlims}
In this paper, we will be concerned with scattering amplitudes involving $n$ massless gravitons and a single massive scalar. We will always label the graviton momenta by $p_1, \dots, p_n$ and the scalar momentum will be denoted by $p_{n+1}$. In order to compute the corresponding celestial amplitudes, it is convenient to use the following parametrization for the massless momenta:
\begin{align}
    \label{eq:masslessmompar22}
    p_i = \e_i \omega_i \left(1+z_i \zb_i, z_i+\zb_i, z_i - \zb_i, 1-z_i \zb_i \right), \qquad i = 1, \dots, n~. 
\end{align}
This parametrization is written for a spacetime with $(2,2)$ signature (specifically $(+,-,+,-)$). $\om_i$ is a positive real number while $z_i, \zb_i$ are real and independent. $\e_i = \pm 1$ is the direction of the momentum. The parametrization for $(3,1)$ signature is obtained by Wick rotating the third component. The scattering amplitude will be written in terms of spinor helicity variables, which for the momentum parametrization above, take the form
\begin{align}
    \label{eq:shvars}
    \lambda_i = \sqrt{2 \omega_i} \e_i\begin{pmatrix}
        1\\ z_i
    \end{pmatrix}, \qquad  \tilde{\lambda}_i = \sqrt{2 \omega_i}\begin{pmatrix}
        1\\ \zb_i
    \end{pmatrix}, \qquad i = 1, \dots, n~.
\end{align} 
The analogous parametrization for the massive scalar momentum (also in $(2,2)$ signature) is 
\begin{align}
\label{eq:massivemompar22}
    p_{n+1} = \frac{m \e_{n+1}}{2y} \left( 1+y^2 + z_{n+1} \zb_{n+1} , z_{n+1}+\zb_{n+1}, z_{n+1}-\zb_{n+1}, 1-y^2-z_{n+1} \zb_{n+1}\right)~.
\end{align}
Here $\e_{n+1} = \pm 1$ is the direction of the momentum, $y, z_{n+1}, \zb_{n+1}$ are real and independent with $y>0$. The celestial amplitude can be obtained from the momentum space scattering amplitude by changing the basis from plane waves to conformal primary wavefunctions. For the case of $n$ massless and one massive particle, this is implemented by \cite{Pasterski:2016qvg, Pasterski:2017ylz} 
\begin{multline}
\label{eq:celampdef}
    \tilde{\mA}_{n+1}^{\e} \left(\lbrace \D_1, z_1, \zb_1\rbrace^{J_1}, \dots, \lbrace \D_n, z_n, \zb_n \rbrace^{J_n} ;\lbrace \D_{n+1}, w, \bar{w}\rbrace\right)   = \\ 
    \frac{m^2}{4}\int_0^{\infty} \prod_{i=1}^n \frac{d\omega_i}{\omega_i} \omega_i^{\D_i}\int_0^{\infty} \frac{dy}{y^3} \int_{-\infty}^{\infty} dz_{n+1} \, d\zb_{n+1} \\
    G_{\D_{n+1}} \left(y, z_{n+1},\zb_{n+1};w,\bar{w}\right)  \mA_{n+1}\left(p_1^{J_1}, \dots, p_n^{J_n}, p_{n+1} \right)~.
\end{multline}
Here $J_1, \dots, J_n$ are the helicities of the massless particles and 
\begin{align}
    G_{\D} \left(y, z_{n+1},\zb_{n+1};w,\bar{w}\right) = 
    \left(\frac{y}{y^2+|z_{n+1}-w|^2}\right)^{\D}
\end{align}
is the bulk-to-boundary propagator. The superscript $\e$ in (\ref{eq:celampdef}) is meant to indicate the dependence of the celestial amplitude on the directions of the momenta. The quantity of interest in this paper is\footnote{The double brackets in the LHS of (\ref{eq:mainobject}) is meant to serve as a reminder of the presence of the massive scalar profile and we use $\mAt$ for the celestial amplitude with the conformal dimension of the scalar set to 0.} 
\begin{align}
\label{eq:mainobject}
     \tilde{\mathrm{A}}_n &= \left\langle\langle \mathcal{O}_{\D_1, J_1}\left(z_1, \zb_1\right) \dots \mathcal{O}_{\D_n, J_n}\left(z_n, \zb_n\right) \right\rangle\rangle \\
     &\nonumber\equiv \sum_{\e_i}\tilde{\mA}_{n+1}^{\e}
     \left(\left\lbrace \D_1, z_1, \zb_1 \right\rbrace^{J_1}, \dots, \left\lbrace \D_n, z_n, \zb_n \right\rbrace^{J_n}; \left\lbrace 0, w, \bar{w} \right\rbrace \right)
\end{align}
whose features we now explain. But first, we pause to make a comment on notation. We will use $\an{\an{\dots}}$ and $\tilde{\mathrm{A}}_n$ interchangeably --- the former being used more often for brevity and the latter when we wish to emphasize its nature as a CCFT correlator. Implicit in both is the fact that we have set the conformal dimension of the scalar, $\Delta_{n+1}$  to 0. The subscript on $\tilde{\mathrm{A}}_n$ matches the number of operators in the double brackets $\an{\an{\dots}}$ and is one less than the subscript on the corresponding momentum space scattering amplitude. Returning to the features of $\tilde{\mathrm{A}}_n$, firstly note that setting $\D_{n+1} = 0$ eliminates all dependence on the scalar coordinates $w, \bar{w}$. The RHS of (\ref{eq:mainobject}) behaves like an $n$-point correlator\footnote{In the celestial amplitude wherein the conformal dimension of the scalar has been set to zero, the subscript $n$ of $\tilde{\mA}_n$ counts only the number of gravitons.} --- a claim that shall be supported by the computations in this paper. When $\phi$ is a non-dynamical field, this has the interpretation of being the Mellin transform of an amplitude on a scalar background and has been studied for gluons in \cite{Casali:2022fro}. These amplitudes take on a particularly simple form \cite{Dixon:2004za} due to the absence of scalar exchange contributions to the amplitude. Upon including gravity, we must promote $\phi$ to a dynamical field and include the resulting scalar exchange terms in the amplitude.\\

Secondly, despite the absence of manifest Lorentz invariance due to the presence of a three-dimensional integral, (\ref{eq:mainobject}) is indeed Lorentz invariant. The three-dimensional measure corresponds to integration over the Lorentz invariant phase space of one massive particle:
\begin{align}
    \int \frac{d^4Q}{\left(2\pi\right)^4} \, \delta^{+} \left(Q^2-m^2\right)  = \int \frac{d^3\vec{Q}}
    {\left(2\pi\right)^3 2 Q^0} = \frac{m^2}{4}\int_0^{\infty} \frac{dy}{y^3}\, \int_{-\infty}^{\infty} dz\, d\zb~.
\end{align}
Here $Q$ is an off-shell momentum which is put on-shell (with positive energy) by the $\delta$ function. Once this is on-shell, we can use a parametrization similar to (\ref{eq:massivemompar22}) and perform a change of variables as shown in the second equality above. \\

Finally, we sum over the directions $\e_i$ purely for practical purposes since this simplifies computations, particularly in Klein space. This simplification has already been exploited before \cite{Fan:2021isc,Hu:2022syq,De:2022gjn}.\footnote{This procedure computes the celestial amplitude for certain ``boost + $\mathbb{Z}_2$" eigenstates \cite{Jorge-Diaz:2022dmy}.} 

\section{Massive scalar-graviton scattering amplitudes} 
\label{sec:scamps}
We must first construct the scattering amplitudes in a theory of scalars and gravitons. We will make a choice for the three-point amplitudes motivated by the results of \cite{Casali:2022fro}. In this work, the authors considered gluons coupled to a single massive, uncolored scalar which resulted in the following three-point amplitudes
\begin{align}
\label{eq:YM3ptamps}
    \mA_3 \left(1^+, 2^+, 3^-\right) = \kappa_{1,1,-1} \frac{\sq{12}^3}{\sq{23}\sq{31}}, \qquad  \mA_3 (1^{+},2^{+},3^{\phi}) &= \kappa_{1,1,0} \sq{12}^2~.
\end{align}
Here and in the rest of this paper, we will leave the momentum conserving delta function implicit while writing scattering amplitudes in momentum space. The authors of \cite{Casali:2022fro} worked in $(2,2)$ signature and this corresponded to introducing terms proportional to $\phi \, \text{Tr} \left(F_+\right)^2$  in the Lagrangian where
\begin{align}
    F_{+}^{a\, \mu \nu} = \frac{1}{2}\left(F^{a \,\mu \nu} + \frac{1}{2} \epsilon^{\mu \nu \rho \sigma} F^a_{\rho \sigma}\right)
    \label{eq:fieldstrengths}
\end{align}
is the self-dual gauge field strength.\footnote{This is real in (2,2) signature.} $\kappa_{1,1,-1}$ is the Yang-Mills coupling and $\kappa_{1,1,0}$ is related by some numerical factor to the Wilson coefficient of the non-minimal $\phi \, \text{Tr} \left(F_+\right)^2$ interaction. The scalar is uncolored and does not couple minimally to the gluon.\\

The analogous three-point amplitudes in gravity can be easily written down and take the following form 
\begin{align}
    \label{eq:grav3ptamps1}
    \mA_3 \left(1^{++}, 2^{++}, 3^{--}\right) = \kappa_{2,2,-2} \frac{\sq{12}^6}{\sq{23}^2\sq{13}^2}~, \qquad  \mA_3 \left(1^{++}, 2^{++}, 3^{\phi}\right) &= \kappa_{2,2,0} \,\sq{12}^4~. 
\end{align}
However, it is also necessary to introduce the scalar-graviton minimal coupling \cite{Arkani-Hamed:2017jhn}
\begin{align}
    \label{eq:grav3ptamps2}
    \mA_3 \left(1^{++}, 2^{\phi}, 3^{\phi}\right) &= \kappa_{2,0,0} \frac{\left\langle \xi_1| p_{2} |1 \right]}{\an{\xi_1 1}} \frac{\left\langle \xi_2 | p_{2} |1 \right]}{\an{\xi_2 1}}~,
\end{align}
where $\xi_1, \xi_2$ indicate arbitrary reference spinors. From the Lagrangian perspective, the minimal coupling is unavoidable in order to maintain diffeomorphism invariance. From a purely amplitudes perspective, as we will show later, any four-point amplitude constructed without the minimal coupling (\ref{eq:grav3ptamps2}) is inconsistent with the soft graviton theorem. Finally, we note that the amplitudes in (\ref{eq:grav3ptamps1}, \ref{eq:grav3ptamps2}) arise from terms which are schematically of the form $R^2 \phi, R \phi^2$ in the Lagrangian. \\

These three-point amplitudes serve as seeds for the construction of four- and higher-point ones. We utilize recursion relations, in particular those generated by a three-line shift \cite{Risager:2005vk} for their construction, which begins with defining
\begin{align}
\label{eq:3lineshift}
    \hat{\lambda}_i = \lambda_i - z w_i X~, \qquad i = 1,2,3
\end{align}
such that $w_i \tilde{\lambda}_i = 0$ and $X$ is an arbitrary reference spinor which must drop out at the end. A particularly convenient choice is
\begin{align}
    w_1 = \sq{23}~, \quad w_2 = \sq{31}~, \quad w_3 = \sq{12}~.
\end{align}
The three shifted momenta are now complex and functions of $z$. Locality dictates that poles can occur at
\begin{align}
 z_{\star} \quad \text{such that} \quad \hat{P}^2 = 0 \quad \text{or} \quad \hat{P}^2 = m^2~,
\end{align}
where $\hat{P}$ is a sum of external momenta necessarily involving at most two of $\hat{p}_1, \hat{p}_2, \hat{p_3}$. The residues on these poles are governed by unitarity implying that they factorize into a product of lower point amplitudes. The amplitude can thus be written as (assuming that there is no contribution from the pole at infinity)
\begin{align}
   \mA_n \equiv  \mA_n \left(0\right) = \sum_{z_{\star}} \left[ \mA_m^{(L)}\left(z_{\star}\right) \frac{1}{P^2}  \mA_{n-m}^{(R)}\left(z_{\star}\right) + \mA_m^{(L)}\left(z_{\star}\right) \frac{1}{P^2-m^2}  \mA_{n-m}^{(R)} \left(z_{\star}\right)\right]~.
\end{align}
We will now employ this to construct the four- and five-point graviton-scalar amplitudes.
\subsection{Four-point amplitudes}
We first consider the amplitude $\mA \left(1^{++}, 2^{++}, 3^{++}, 4^{\phi}\right)$ which receives contributions from both scalar and graviton exchange channels
\begin{equation}
    \mA_4 \left(1^{++}, 2^{++}, 3^{++}, 4^{\phi} \right) \equiv \mA_4^{gr,s} + \mA_4^{gr,t} + \mA_4^{gr, u} + \mA_4^{sc,s} + \mA_4^{sc,t} + \mA_4^{sc,u}.
\end{equation}
The first of these terms corresponds to an $s$-channel graviton exchange 
\begin{align}
\label{eq:4ptgravterm1-1}
    \mA_4^{gr, s} &= \vcenter{\hbox{\includegraphics[scale=0.6]{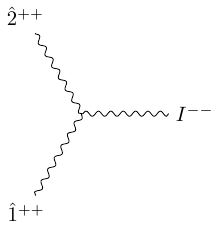}
     \includegraphics[scale=0.6]{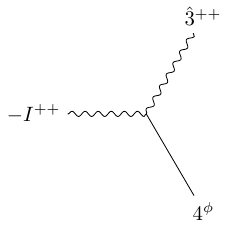}}}  \\
  &\nonumber=\mA_3 \left(\hat{1}, \hat{2}, P \right) \frac{1}{P^2} \mA_3 \left(\hat{3}, 4, -P\right)~,
\end{align}
with the value of $z_{\star}$ given by
\begin{align}
    \left(\hat{p}_1 + \hat{p}_2\right)^2 = 0 \implies z_{\star} = \frac{\an{12}}{w_1 \an{X2}-w_2 \an{X1}}~.
\end{align}
Using the three-point amplitudes in (\ref{eq:grav3ptamps1}, \ref{eq:grav3ptamps2}), this evaluates to
\begin{align}
\label{eq:4ptgravterm1-2}
    \mA_4^{gr, s} &= \kappa_{2,2,-2}\left(\frac{\sq{12}^3}{\sq{2I}\sq{I1}}\right)^2 \frac{1}{\an{12}\sq{21}}\kappa_{2,2,0} \sq{3I}^4\\
    &\nonumber = -\kappa_{2,2,-2}\kappa_{2,2,0} \frac{\sq{12}}{\an{12}}\frac{\left[3|p_4|X\right\rangle^4}{\an{1X}^2\an{2X}^2}~.
\end{align}
Similarly, the $s$-channel scalar exchange term is
\begin{align}
\label{eq:4ptscalarterm1}
    \mA_4^{sc, s} &= \vcenter{\hbox{\includegraphics[scale=0.6]{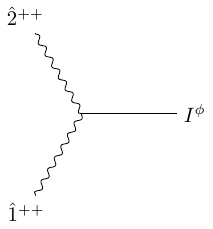}
     \includegraphics[scale=0.6]{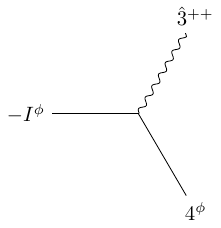}}}\\
    &\nonumber= \mA_3 \left(\hat{1}, \hat{2}, P \right) \frac{1}{P^2-m^2} \mA_3 \left(\hat{3}, 4, -P\right)\\
   &\nonumber =\kappa_{2,2,0} \sq{12}^4 \frac{1}{s - m^2}  \kappa_{2,0,0}\frac{\left[3|p_4|\xi_1\right\rangle}{\an{\hat{3}\xi_1}}\frac{\left[3|p_4|\xi_2\right\rangle}{\an{\hat{3}\xi_2}}~.
\end{align}
Here $s = \left(p_1+p_2\right)^2$ is the usual Mandelstam invariant. The remaining terms can be obtained by the replacements $1 \leftrightarrow 3$ and  $2 \leftrightarrow 3$. Adding all of these terms and setting the arbitrary reference spinors $\xi_1 = \xi_2 = X$, we finally get
\begin{multline}
    \mA_4 \left(1^{++}, 2^{++}, 3^{++}, 4^{\phi} \right) = -\kappa_{2,2,-2}\kappa_{2,2,0} \frac{\sq{12}}{\an{12}}\frac{\left[3|p_4|X\right\rangle^4}{\an{1X}^2\an{2X}^2} +\kappa_{2,0,0}\kappa_{2,2,0} \frac{\sq{12}^4 }{s - m^2} \frac{\left[3|p_4|X\right\rangle^2}{\an{3X}^2}\\
    \qquad\qquad + \qquad\qquad  (2 \leftrightarrow 3) \qquad\qquad + \qquad\qquad (1 \leftrightarrow 3)~.\nonumber 
\end{multline}
The spinor $X$ is arbitrary and the amplitude must be independent of it. It can be checked that the dependence on the arbitrary spinor $X$ drops out {\it only} if $\kappa_{2,0,0} = \kappa_{2,2,-2}$ and in that case the expression simplifies to
\begin{multline}
     \label{eq:4ptgravscalar}
    \mA_4 \left(1^{++}, 2^{++}, 3^{++}, 4^{\phi} \right) = \kappa_{2,2,-2}\kappa_{2,2,0}m^4 \frac{\sq{12}}{\an{12}}\frac{\sq{23}}{\an{23}}\frac{\sq{31}}{\an{31}}\\ 
    \times \left(-2 + \frac{s}{s-m^2} + \frac{t}{t-m^2} + \frac{u}{u-m^2}\right),
\end{multline}
with $t= \left(p_1+p_4\right)^2$ and $u = \left(p_1+p_3\right)^2$. We pause here to point out that this expression involves a nontrivial cancellation of the $X$ dependence between the graviton exchange terms (which arise from the non-minimal coupling $\kappa_{2,2,0}$) and the scalar exchange terms (arising from the minimal coupling $\kappa_{2,0,0}$).  This affirms the statement made in the beginning of this section that the presence of this minimal coupling is crucial to have a consistent and non-zero four-point amplitude. Furthermore, it is also necessary to have $\kappa_{2,0,0} = \kappa_{2,2,-2}$ --- a relation which is to be expected on grounds of the equivalence principle. It is interesting to compare this to the four-point amplitude involving one scalar and three positive helicity gluons originally computed in \cite{Dixon:2004za} to be 
\begin{align}
    \mathcal{A}_4(1^{+},2^{+},3^{+},4^{\phi}) = \kappa_{1,1,-1} \kappa_{1,1,0}\,\frac{m^4}{\an{12} \an{23} \an{13}}~.
\label{allplusphi4pt}
\end{align}
This amplitude is particularly simple since the absence of the gluon-scalar minimal coupling eliminates massive exchange contributions.\\

Moving on to the other four-point amplitudes in this theory, $\mA \left(1^{++}, 2^{++}, 3^{\phi}, 4^{\phi} \right)$ has been computed in \cite{Britto:2021pud} via BCFW recursion relations to be 
\begin{align}
    \label{eq:2grav2scalar}
     \mA_4 \left(1^{++}, 2^{++}, 3^{\phi}, 4^{\phi} \right) &= \kappa^2_{2,2,-2} m^4 \frac{\sq{12}^4}{s \left(t-m^2\right) \left(u-m^2\right)}~.
\end{align}
Finally, the amplitude $\mA \left(1^{++}, 2^{++}, 3^{--}, 4^{\phi}\right)$ cannot be computed by a three-line shift (or using BCFW recursion). We provide an independent derivation of this amplitude in Appendix \ref{app:++-phi} and show that it is given by
\begin{align}
\label{eq:gravppms}
    \mA_4 \left(1^{++}, 2^{++}, 3^{--}, 4^{\phi} \right) = \kappa_{2,2,0}\kappa_{2,-2,-2} \frac{\sq{12}^6 \an{23}^2\an{13}^2}{\left(s-m^2\right)t u}~.
\end{align}

\subsection{Five-point amplitudes}
The five-point computation proceeds in a similar, albeit more complicated manner to that at four points. The three-line shift in (\ref{eq:3lineshift}) and the associated recursion relation imply that the amplitude $\mA\left(1^{++},2^{++}, 3^{++}, 4^{++}, 5^{\phi}\right)$ receives contributions from 18 compatible factorization channels.
\begin{equation}
\label{eq:5ptgravscalar}
     \mA_5 \left(1^{++}, 2^{++}, 3^{++}, 4^{++}, 5^{\phi}\right) = \sum_{i=1}^{3}\sum_{j=i+1}^{4}\left(\mA_5^{gr, i,j} +\mA_5^{sc, i,j}\right) + \sum_{i=1}^3 \mA_5^{sc,i,5}.
\end{equation}
They can be divided into five classes with $\mA_5^{gr,1,2},\mA_5^{gr,3,4},\mA_5^{sc,1,2},\mA_5^{sc,3,4},\mA_5^{gr,1,5}$ being their representatives. The rest can be obtained by making appropriate replacements on the external legs.  As we will now demonstrate, it will be useful to set $X = \lambda_4$.

\begin{center}
\underline{\bf Evaluation of $\mA_5^{gr,3,4}$ and $\mA_5^{sc,3,4}$}
\end{center}
These classes of terms arise from graviton and scalar exchanges and correspond to the momentum $\left(\hat{p}_1+p_4\right)^2$ going on-shell. 
\begin{align}
\label{eq:gr14}
    \mA_5^{gr, 3,4} &= \vcenter{\hbox{\includegraphics[scale=0.8]{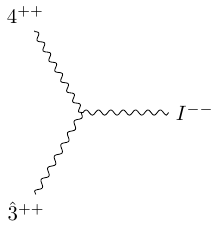}
     \includegraphics[scale=0.8]{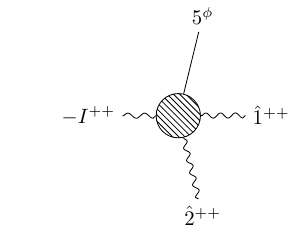}}}\\
     &=\mA_3 \left(\hat{3}^{++}, 4^{++}, I^{--}\right) \frac{1}{P^2} \mA_4\left(-I^{++}, \hat{2}^{++}, \hat{1}^{++}, 5^{\phi}\right)\nonumber\\
      &\nonumber = \left(\frac{\sq{34}^3}{\sq{4I}\sq{I3}}\right)^2 \frac{1}{\an{34}\sq{43}} \frac{\sq{12}}{\an{\hat{1}\hat{2}}}\frac{\sq{2I}}{\an{\hat{2}I}}\frac{\sq{I1}}{\an{I\hat{1}}}\\
      &\nonumber \qquad\qquad \times \left( -2 + \frac{s_{\hat{1}\hat{2}}}{s_{\hat{1}\hat{2}}-m^2} + \frac{s_{\hat{2}\hat{3}4}}{s_{\hat{2}\hat{3}4}-m^2} + \frac{s_{\hat{1}\hat{3}4}}{s_{\hat{1}\hat{3}4}-m^2}\right)\nonumber~,
\end{align}
where we've used the notation $s_{ij} = \left(p_i+p_j\right)^2, s_{ijk} = \left(p_i+p_j+p_k\right)^2$. After some manipulation this term takes the form  
\begin{align}
\label{eq:amp5vanishingterm1}
    \mA_5^{gr,3,4} 
    &= -\frac{\sq{34}}{\an{34}} \frac{\sq{12}}{\an{\hat{1}\hat{2}}} \frac{\left[2|\hat{3}+4|X\right\rangle \left[1|\hat{3}+4|X\right\rangle}{\an{3X}\an{4X}\an{\hat{1}\hat{3}}\an{\hat{2}4}} \left[ -2 + \frac{s_{\hat{1}\hat{2}}}{s_{\hat{1}\hat{2}}-m^2} + \frac{s_{\hat{2}\hat{3}4}}{s_{\hat{2}\hat{3}4}-m^2} + \frac{s_{\hat{1}\hat{3}4}}{s_{\hat{1}\hat{3}4}-m^2}\right]~.
\end{align}
The shifted spinors are determined by first solving $\an{\hat{3}4} = 0$ for $z$. This gives
\begin{align}
    z = \frac{\an{34}}{\an{X4}}~,\quad \hat{\lambda}_1 = \lambda_1 - \frac{\sq{23}\an{34}}{\sq{12}\an{X4}}X~, \quad \hat{\lambda}_2 = \lambda_2 - \frac{\sq{31}\an{34}}{\sq{12}\an{X4}}X~, \quad \hat{\lambda}_3 = \frac{\an{X3}}{\an{X4}} \lambda_4~.
\end{align}
Plugging all of this into (\ref{eq:amp5vanishingterm1}), we find that 
\begin{align}
   \mA_5^{gr, 3,4} \propto \an{X4}~,
\end{align}
which vanishes when $X=\lambda_4$. A similar analysis for the scalar exchange diagrams shows that it also vanishes. Thus all diagrams in this class do not contribute to the amplitude and we have
\begin{align}
\label{eq:amp5vanishingterms}
    \mA_5^{gr, 1,4} = \mA_5^{gr, 2,4} = \mA_5^{gr, 3,4} =  \mA_5^{sc, 1,4} = \mA_5^{sc, 2,4} = \mA_5^{sc, 3,4} = 0~.
\end{align} 

\vspace{3pt}

\begin{center}
    {\underline{\bf Evaluation of $\mA_5^{gr, 1,2}$}}
\end{center}
This class of terms arise from the momentum $\left(\hat{p}_1+\hat{p}_2\right)^2$ going on-shell and corresponds to the exchange of a graviton. This channel cannot be obtained from (\ref{eq:gr14}) because the shift does not treat $p_4$ on the same footing as $p_1, p_2, p_3$.
\begin{align}
\label{eq:gr12}
    \mA_5^{gr, 1,2} &= \vcenter{\hbox{\includegraphics[scale=0.8]{Figs/4ptamp1part1.pdf}
     \includegraphics[scale=0.8]{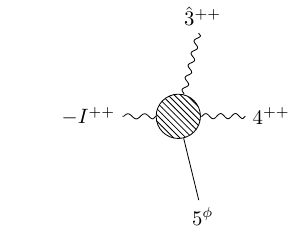}}}\\
     &=\mA_3 \left(\hat{1}^{++}, \hat{2}^{++}, I^{--}\right) \frac{1}{P^2} \mA_4\left(-I^{++}, \hat{3}^{++}, 4^{++}, 5^{\phi}\right)\nonumber\\
     &\nonumber=\left(\frac{\sq{12}^3}{\sq{2I}\sq{I1}}\right)^2 \frac{1}{\an{12}\sq{21}}\frac{\sq{34}\sq{4I}\sq{I3}}{\an{\hat{3}\hat{4}}\an{4I}\an{I\hat{3}}}\\
     &\nonumber \qquad\qquad \times \left(-2 + \frac{s_{\hat{3}4}}{s_{\hat{3}4}-m^2} + \frac{s_{\hat{1}\hat{2}4}}{s_{\hat{1}\hat{2}4}-m^2} + \frac{s_{\hat{1}\hat{2}\hat{3}}}{s_{\hat{1}\hat{2}\hat{3}}-m^2}\right)~.
\end{align}
The kinematic invariants involving shifted spinors can be evaluated by first determining the value of $z$ for this channel by setting $\an{\hat{1}\hat{2}} = 0$ and plugging in this value of $z$ into (\ref{eq:3lineshift}). On doing this, we get
\begin{align}
   &\an{\hat{3}4}=\an{34}, \qquad \an{4I} = -1, \qquad \an{I\hat{3}} = -\frac{s_{123}}{\left[3|p_1+p_2|4\right\rangle}\\
   &s_{\hat{1}\hat{2}4} = s_{124}-s_{12}, \qquad s_{\hat{1}\hat{2}\hat{3}} = s_{123}~.\nonumber
\end{align}
This brings this term to the form
\begin{multline}
    \mA_5^{gr, 1,2} = -\frac{\sq{12}}{\an{12}}\frac{\sq{34}}{\an{34}}\frac{\sq{14}}{\an{14}}\frac{\sq{24}}{\an{24}} \left(\frac{1}{s_{24}} + \frac{1}{s_{14}}\right) \frac{\left[3|p_5|4\right\rangle^2}{s_{123}}\\
    \times \left( -2 + \frac{s_{34}}{s_{34}-m^2} + \frac{s_{14}+s_{24}}{s_{14}+s_{24}-m^2} + \frac{s_{123}}{s_{123}-m^2}\right)~.
\end{multline} 

\newpage

\begin{center}
    \underline{\bf Evaluation of $\mA_5^{sc, 1,5}$}
\end{center}
This class of terms arises from the momentum $\left(\hat{p}_1+ p_5\right)^2$ going on-shell and corresponds to a scalar exchange. 
\begin{align}
\label{eq:sc15}
     \mA_5^{sc, 1,5} &= \vcenter{\hbox{\includegraphics[scale=0.8]{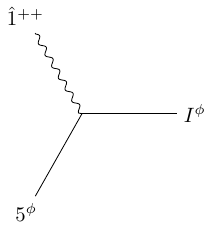}
     \includegraphics[scale=0.8]{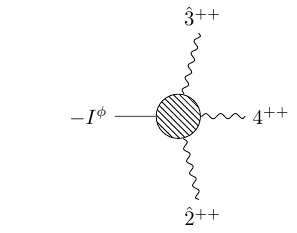}}}\\
     &=\mA_3 \left(\hat{1}^{++}, 5^{\phi}, I^{\phi}\right) \frac{1}{P^2-m^2} \mA_4\left(-I^{\phi}, \hat{3}^{++}, 4^{++}, \hat{2}^{++}\right)\nonumber\\
      &\nonumber = \frac{\left[1|p_5|\xi_1\right\rangle\left[1|p_5|\xi_2\right\rangle}{\an{\hat{1}\xi_1}\an{\hat{2}\xi_2}} \frac{1}{s_{15}-m^2} \frac{\sq{23}\sq{34}\sq{42}}{\an{\hat{2}\hat{3}}\an{\hat{3}4}\an{4\hat{2}}}\\
     &\nonumber \qquad\qquad \times \left( -2 + \frac{s_{\hat{3}4}}{s_{\hat{3}4}-m^2} +\frac{s_{\hat{3}\hat{2}}}{s_{\hat{3}\hat{2}}-m^2}+ \frac{s_{\hat{2}4}}{s_{\hat{2}4}-m^2} \right)~.
\end{align}
In this case, the kinematic invariants involving shifted momenta evaluate to 
\begin{align}
    \an{\hat{2}\hat{3}} = \frac{s_{15}-s_{23}-m^2}{\sq{23}}~, \qquad s_{\hat{2}\hat{3}} = s_{23}-s_{15}+m^2~, \qquad s_{\hat{3}4} = s_{34}~, \qquad s_{\hat{2}4} = s_{24}~.
\end{align}
Setting $\xi_1 = \xi_2 = \lambda_4$ for simplicity, we get
\begin{align}
    \nonumber \mA_5^{sc,1,2} =& \frac{\sq{23}}{\an{23}}\frac{\sq{34}}{\an{34}}\frac{\sq{24}}{\an{24}}\frac{\sq{14}}{\an{14}} \frac{s_{23}}{s_{14}}\frac{\left[1|p_5|4\right\rangle^2}{\left(s_{15}-m^2\right) \left(s_{15}-s_{23}-m^2\right)}\\
    &\nonumber \qquad\qquad \times \left( -2 + \frac{s_{34}}{s_{34}-m^2} + \frac{s_{123}+s_{14}}{s_{123}+s_{14}-m^2} + \frac{s_{24}}{s_{24}-m^2}\right).
\end{align}

\vspace{3pt}
  
\begin{center}
    \underline{\bf Evaluation of $\mA_5^{sc, 1,2}$}
\end{center} 
This final class of terms also corresponds to a scalar exchange but arises from the momentum $\left(\hat{p}_1+\hat{p}_2\right)^2$ going on-shell. 
\begin{align}
\label{eq:sc12}
    \mA_5^{sc, 1,2} &= \vcenter{\hbox{\includegraphics[scale=0.8]{Figs/4ptamp2part1.pdf}
     \includegraphics[scale=0.8]{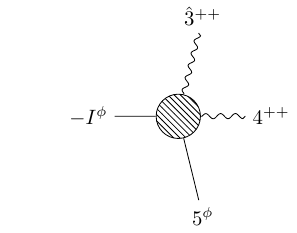}}}\\
     &=\mA_3 \left(\hat{1}^{++}, \hat{2}^{++}, I^{\phi}\right) \frac{1}{P^2-m^2} \mA_4\left(-I^{\phi}, \hat{3}^{++}, 4^{++}, 5^{\phi}\right)\nonumber\\
     &\nonumber = \sq{12}^4\, \frac{1}{s_{12}-m^2} \,\frac{\sq{34}^4}{s_{\hat{3}4}\left(s_{\hat{1}\hat{2}4}-m^2\right)\left(s_{\hat{1}\hat{2}\hat{3}}-m^2\right)}~.
\end{align}
Here the shifted momenta are determined by $\an{\hat{1}\hat{2}}\sq{12}=m^2$ and this term evaluates to 
\begin{equation}
    \mA_5^{sc, 1,2} = \frac{\sq{12}^4 \sq{34}^4}{\left(s_{12}-m^2\right)s_{34}\left(s_{123}-m^2\right) \left(s_{14}+s_{24}\right)}~.
\end{equation} 
This completes the computation of $\mA_5 \left(1^{++}, 2^{++}, 3^{++}, 4^{++}, 5^{\phi} \right)$. In computing this amplitude using the recursion relations, we have implicitly assumed that there is no contribution from a pole at infinity in the complex $z$ plane. In order to verify the validity of this assumption, we have checked that the amplitude we computed has all the correct soft and collinear limits.

\section{Celestial amplitudes}
\label{sec:celestialscalargravamp}
We are interested in evaluating the celestial amplitudes corresponding to the bulk scattering amplitudes of the previous section. As explained in Section \ref{sec:perlims}, we are interested in the correlators obtained by setting $\Delta =0$, where $\Delta$ is the conformal dimension of the massive scalar. Since all the dependence on the scalar coordinates drops out of the correlation functions, we will refer to the celestial counterparts of $n+1$-point bulk scattering amplitudes as $n$-point correlators. This is further supported, as we will now show, by the structure of these correlators. 
\subsection{Two-point function}
The momentum space three-point amplitude (\ref{eq:grav3ptamps1}), written using the parameterization (\ref{eq:masslessmompar22}, \ref{eq:massivemompar22}) is given by
\begin{align}
    \mA_3 \left(1^{++}, 2^{++}, 3^{\phi} \right) = \kappa_{2,2,0} \,m^4\, \frac{\zb_{12}^2}{z_{12}^2} \delta^{(4)} \left(p_1 + p_2 + Q\right)~.
\end{align}
We have made use of momentum conservation ($\an{12}\sq{12}=m^2$) in arriving at the above form. The corresponding celestial amplitude integrated over the scalar phase space is given by the expression 
\begin{align}
\label{eq:2ptmomspace}
    &\mAt_2 \left(\left\lbrace \D_1,z_1,  \zb_1\right \rbrace^{++}, \left\lbrace \D_2, z_2, \zb_2\right\rbrace^{++}\right) \nonumber \\
    &= \kappa_{2,2,0} \, \sum_{\epsilon_1, \e_3} \int_0^{\infty} dy \, dz_3 \, d\zb_3 \, \frac{m^2}{4y^3} 
    \times\int_0^{\infty} d\omega_1 \,d\omega_2 \,\omega_1^{\D_1-1} \, \omega_2^{\D_2-1} \, \frac{\zb_{12}^2}{z_{12}^2} \, \delta^{(4)} \left(p_1 + p_2 + Q\right)~.
\end{align}
We can use the $\delta$-function to solve for $y, z, \zb, \omega_1$ to get 
\begin{align}
    \label{eq:deltafunc}
    \delta^{(4)} \left(p_1+p_2+Q\right) = \frac{1}{\left|J\right|}\delta \left(y-y^{\star}\right)\delta \left(\zb_3-\zb^{\star}_3\right)\delta \left(z_3-z_3^{\star}\right) \delta \left(\omega_1 - \omega_1^{\star}\right)~,
\end{align}
where\footnote{We have not displayed the solutions for $z_3,\zb_3$ since they are not required for this computation.}
\begin{align}
   \omega_1^{\star} = \frac{m^2}{4\e_1 \e_2 z_{12}\zb_{12} \omega_2}, \qquad y^{\star} = \frac{-2m \e_2 \e_3 z_{12} \zb_{12}\omega_2}{m^2 + 4 z_{12} \zb_{12} \omega_2^2}, \qquad J = \frac{m^2 z_{12}\zb_{12}\e_1 \e_2 \om_2}{(y^{\star})^3}~.
\end{align}
After integrating over $y, z_3, \zb_3, \om_1$, we get 
\begin{equation}
     \mAt_2 = \kappa_{2,2,0} m^2 \left(\frac{\zb_{12}}{z_{12}}\right)^2 \sum_{\e_1,\e_3} \int_0^{\infty} d\omega_2 \, \left(\frac{m^2}{4\e_1 \e_2 z_{12}\zb_{12} \omega_2}\right)^{\D_1} \, \omega_2^{\D_2-1} \,\Theta \left(\omega_1^{\star}\right) \, \Theta \left(y^{\star} \right)~.
\end{equation}
The integral over the $\delta$-function gives rise to the two $\Theta$ functions since the $y, \omega_1$ integrals are over $\mathbb{R}^+$. The sum over $\e_3$ eliminates $\Theta \left(y^{\star} \right)$ while the sum over $\e_1$ eliminates $\Theta \left(\omega_1^{\star}\right)$ yielding 
\begin{align}
      \mAt_2  &= \kappa_{2,2,0} m^2 \left| \frac{\zb_{12}}{z_{12}}\right|^2 \left|\frac{m^2}{4z_{12}\zb_{12}}\right|^{\D_1} \int_0^{\infty} d\omega_2 \, \omega_2^{\D_2-\D_1-1}~.
\end{align}
Performing the $\omega_2$ integral, we obtain the final result for the two-point celestial correlator
\begin{equation}
\an{\an{\mathcal{O}^{++}_{\D_1}\left(z_1,\zb_1\right)\mathcal{O}^{++}_{\D_2}\left(z_2,\zb_2\right)}} \equiv \mAt_2 = \kappa_{2,2,0} m^2 \left|\frac{\zb_{12}}{z_{12}}\right|^2 \left|\frac{m^2}{4z_{12}\zb_{12}}\right|^{\D_1} \times 2 \pi \delta\left(\D_1 - \D_2\right)~.
\label{two-pointcelestamp}
\end{equation}
We see that the resulting object indeed resembles a two-point correlation function as in an ordinary CFT. \\

It is also natural to consider correlation functions like {\small $\an{\mo_{\D_1}^{++}\left(z_1, \zb_1\right) \mo_{\D_2}^{0}\left(z_2, \zb_2\right) \mo_{0}^{0}\left(z_3, \zb_3\right)}$}. Interpreting it as a two-point function of a graviton and scalar, we expect it to vanish from the 2D CFT perspective. However, it is nonzero and is suggestive of the fact that additional insertions of the scalar cannot be treated on the same footing as gravitons. We leave the exploration of such correlators and their meaning to future work, noting for now that the singular part of the graviton OPE does not involve the scalar, so the set of correlators $\langle\langle \dots \rangle\rangle$ with only graviton insertions forms a self-consistent sector in some sense, similar in spirit to studying a current algebra independently from any actual CFT.\footnote{We thank Sruthi Narayanan for bringing this to our attention.} 

\subsection{Three-point functions}
\subsubsection{$ \an{\an{\mathcal{O}_{\D_1}^{++} \, \mathcal{O}_{\D_2}^{++} \, \mathcal{O}_{\D_3}^{++}}}$}
The three-point functions are computed from momentum space four-point scattering amplitudes. We start with (\ref{eq:4ptgravscalar}) which we write as
\begin{equation}
    \mA_4 \left(1^{++}, 2^{++}, 3^{++}, 4^{\phi} \right)  = \mA_4^{gr}+\mA_4^{sc,s}+\mA_4^{sc,t}+\mA_4^{sc,u},
    \label{eq:celestial3ptexchanges}
\end{equation}
with $\mA_4^{gr}$ representing the contribution of graviton exchange in all channels. As in the previous section, the $\delta$-function is used to eliminate the scalar momentum variables $z_4,\zb_4, y$ as well as $\omega_3$ 
\begin{align}
    \delta^{(4)} \left(p_1+p_2+p_3+Q\right) =  \frac{1}{\left|J\right|}\delta \left(y-y^{\star}\right)\delta \left(\zb-\zb_{\star}\right)\delta \left(z-z_{\star}\right) \delta \left(\omega_3 - \omega_3^{\star}\right)~,
\end{align}
with 
\begin{align}
    \label{eq:deltasolve4pt}
    &\nonumber\omega_3^{\star} = \frac{\epsilon_3}{4} \frac{m^2 - 4 \omega_1 \omega_2 z_{12} \zb_{12} \epsilon_1 \epsilon_2}{z_{13}\zb_{13} \epsilon_1 \omega_1 + z_{23}\zb_{23} \epsilon_2 \omega_2 }~~,~~ y^{\star} = -\frac{2m \epsilon_4 \left(z_{13}\zb_{13} \epsilon_1 \omega_1 + z_{23}\zb_{23} \epsilon_2 \omega_2\right)}{m^2 + 4 \left(\epsilon_1 \omega_1 z_{13} + \epsilon_2 \omega_2 z_{23} \right)\left(\epsilon_1 \omega_1 \zb_{13} + \epsilon_2 \omega_2 \zb_{23} \right)}~,\\
    &\hspace{5cm} J = \frac{m^2}{y^3}\left(z_{13}\zb_{13} \epsilon_1 \omega_1 + z_{23}\zb_{23} \epsilon_2 \omega_2\right)~.
\end{align}
The resulting three-point celestial correlator for both the graviton and scalar exchanges is given by 
\begin{align}
\label{eq:generic3pt}
    \mAt_3 =\frac{1}{4}\sum_{\e_i} \int_0^{\infty} d\omega_1 \, d\omega_2 \, \omega_1^{\D_1-1} \, \omega_2^{\D_2-1}   \frac{\left(\omega_3^{\star}\right)^{\D_3-1}} 
    {\left|z_{13}\zb_{13} \epsilon_1 \omega_1 + z_{23}\zb_{23} \epsilon_2 \omega_2 \right|} \Theta \left(y^{\star}\right) \Theta\left(\omega_3^{\star}\right) \mA_4. 
\end{align}
We will now compute the above integral for each term in $\mA_4$ (\ref{eq:celestial3ptexchanges}), whose explicit form in the momentum space is given by the relation (\ref{eq:4ptgravscalar}).

\begin{center}
    \underline{\bf Mellin transform of $\mA_4^{gr}$}
\end{center}
In the parametrization of (\ref{eq:masslessmompar22}), the graviton exchange term of the amplitude is 
\begin{align}
\label{eq:tma3step0}
    \mA_4^{gr} = -2\kappa_{2,2,-2}\kappa_{2,2,0}m^4 \frac{\zb_{12}\zb_{23}\zb_{31}}{z_{12}z_{23}z_{31}}.
\end{align}
Plugging this into (\ref{eq:generic3pt}) and performing the sum over $\e_3, \e_4$ gives 
\begin{multline}
\label{eq:magrmellin}
    \mAt_3^{gr} =-2^{1-2\D_3}\kappa_{2,2,-2}\kappa_{2,2,0}m^4 \frac{\zb_{12}\zb_{23}\zb_{31}}{z_{12}z_{23}z_{31}}
    \sum_{\e_1, \e_2} \int_0^{\infty} d\omega_1 \, d\omega_2 \, \omega_1^{\D_1-1} \, \omega_2^{\D_2-1} \\
    \times \left|m^2 - 4 \omega_1 \omega_2 z_{12} \zb_{12} \epsilon_1 \epsilon_2\right|^{\D_3-1}
    \left|z_{13}\zb_{13} \epsilon_1 \omega_1 + z_{23}\zb_{23} \epsilon_2 \omega_2 \right|^{-\D_3}.
\end{multline}
Changing variables to
\begin{align}
    \label{eq:varchange1}
    \omega_1 = \frac{m}{2} \left(\frac{\left|z_{23}\zb_{23}\right| }{\left|z_{12} \zb_{12}\right| \left| z_{13}\zb_{13}\right| }\right)^{\frac{1}{2}} \sqrt{\frac{Y}{X}}~~,~~ \omega_2 = \frac{m}{2} \left(\frac{\left|z_{13}\zb_{13}\right| }{\left|z_{12} \zb_{12}\right| \left| z_{23}\zb_{23}\right| }\right)^{\frac{1}{2}} \sqrt{X Y}~~,~~ \epsilon_1 \epsilon_2 = \epsilon~,
\end{align}
brings the integral to the form 
\begin{align}
\label{eq:tma3step1}
    \mAt_3^{gr} = - \kappa_{2,2,-2}\kappa_{2,2,0}\,\mathcal{Z}\left(z_{ij},\zb_{ij}\right) \, \sum_{\e} f_1 \left(\D_i, \alpha_1 \right) \, f_2 \left(\D_i, \alpha_2\right)~,
\end{align}
with 
\begin{align}
    \mathcal{Z} =  2^{-\beta-3} m^{\beta+5} \frac{\zb_{12} \zb_{23}\zb_{13}}{z_{12}z_{23}z_{13}}\left|z_{12}\zb_{12}\right|^{\frac{\D_3-\D_1-\D_2}{2}} \left|z_{23}\zb_{23}\right|^{\frac{\D_1-\D_2-\D_3}{2}} \left|z_{13}\zb_{13}\right|^{\frac{\D_2-\D_1-\D_3}{2}}~,
\end{align}
being a standard conformal factor and $\beta = -3+\sum_{i=1}^3 \D_i$. $f_1$ and $f_2$ are integrals over the variables $X$ and $Y$ respectively and depend on the $z_{ij},\zb_{ij}$ only through
\begin{equation}
   \alpha_1 =  \epsilon \, \text{sgn} \left(z_{13}\zb_{13}z_{23}\zb_{23}\right),\quad  \alpha_2 = \epsilon \,\text{sgn}\left(z_{12}\zb_{12}\right).
\end{equation}
The $f_1, f_2$ integrals are now readily performed giving us 
\begin{multline}
\label{eq:Xint}
    f_1 \left(\D_i, \alpha_1\right) \equiv \int_0^{\infty} dX \, X^{\frac{\D_2-\D_1+\D_3}{2}-1} \, \left|X +\alpha_1\right|^{-\D_3}\\ =  \frac{\Gamma \left(\frac{\D_2+\D_3-\D_1}{2}\right) \Gamma \left(\frac{\D_1-\D_2+\D_3}{2}\right) }{\Gamma\left(\D_3\right)} 
    \begin{cases}
         1 &\alpha_1 = 1\\
         \frac{\cos \left(\frac{\pi}{2}\left(\D_1 - \D_2\right)\right)}{\cos \frac{\pi}{2} \D_3}   &\alpha_1 = -1~,
    \end{cases}
\end{multline}
and
\begin{multline}
\label{eq:Yint}
  f_2 \left(\D_i, \alpha_2\right) \equiv \int_0^{\infty} dY \, Y^{\frac{\D_1-\D_3+\D_2}{2}-1}  \left|1-\alpha_2 Y \right|^{\D_3-1}   \\
   \qquad =\frac{1}{\pi}  \Gamma \left(\frac{\D_1+\D_2-\D_3}{2}\right) \Gamma \left(\frac{2-\D_1-\D_2-\D_3}{2}\right) \Gamma\left(\D_3\right)\\
   \qquad\qquad\qquad\qquad \times \begin{cases}
        2\sin\frac{\pi}{2} \left(\D_1+\D_2\right) \cos\frac{\pi}{2}\D_3   &\alpha_2 = 1\\
        \sin \pi \D_3 &\alpha_2 = -1~.
    \end{cases}
\end{multline}
Putting all of this together, we finally get two regions for the graviton exchange term
\begin{align}
    \label{eq:Atgr}
    \mAt_3^{gr} = -\kappa_{2,2,-2}\kappa_{2,2,0} \, \mathcal{Z}\left(z_{ij},\zb_{ij}\right)\, \mathcal{G}^{gr}\left(\D_i\right) \,\, \times \,\, \begin{cases} \mathcal{S}_{+}  \qquad &\left(z_{12}\zb_{12}z_{13}\zb_{13}z_{23}\zb_{23}\right)>0\\
        \mathcal{S}_{-} \qquad  &\left(z_{12}\zb_{12}z_{13}\zb_{13}z_{23}\zb_{23}\right)<0~,
    \end{cases}
\end{align}
where
\begin{align}
\label{eq:Ggraviton}
    \mathcal{G}^{gr}\left(\D_i\right) &= \frac{1}{\pi} \, \Gamma \left(\frac{2-\D_1-\D_2-\D_3}{2}\right) \Gamma \left(\frac{\D_1+\D_2-\D_3}{2}\right) \nonumber \\
    & \qquad \qquad \qquad \qquad \qquad \times \Gamma \left(\frac{\D_2+\D_3-\D_1}{2}\right) \Gamma \left(\frac{\D_1-\D_2+\D_3}{2}\right)~,
\end{align}
and
\begin{align}
 \mathcal{S}_{+} &=  \sin \frac{\pi}{2}\left(\D_1+\D_2+\D_3\right) +\sin\frac{\pi}{2}\left(\D_1+\D_2-\D_3\right) \nonumber \\
    & \qquad\qquad\qquad + \sin\frac{\pi}{2}\left(\D_1-\D_2+\D_3\right) + \sin\frac{\pi}{2}\left(\D_2-\D_1+\D_3\right)~, \nonumber \\
    \mathcal{S}_{-} &= \sin\pi \D_1 + \sin\pi \D_2 + \sin\pi \D_3~.
 \label{eq:regionsgrav}
\end{align}

\vspace{2mm}

\begin{center}
    \underline{\bf Mellin transform of $\mA_4^{sc,s},\mA_4^{sc,t}$ and $\mA_4^{sc,u}$}
\end{center}
We only need to evaluate the $s$-channel exchange since the other channels can be simply obtained from this by interchanging $1 \leftrightarrow 3$ and $2 \leftrightarrow 3$. Moreover, since we have the following relation between the graviton and scalar exchange terms
\begin{align}
    \mA_4^{sc,s} = -\frac{1}{2}\mA_4^{gr} \times \frac{4 \e_1 \e_2 \omega_1 \omega_2 z_{12}\zb_{12}}{4 \e_1 \e_2 \omega_1 \omega_2 z_{12}\zb_{12}-m^2}~, 
\end{align}
their Mellin transforms are quite similar. Following the steps that led from (\ref{eq:tma3step0}) to (\ref{eq:tma3step1}), we get 
\begin{align}
     \mAt_3^{sc, s} =\frac{1}{2} \kappa_{2,2,-2}\kappa_{2,2,0} \, \mathcal{Z}\left(z_{ij},\zb_{ij}\right)\sum_{\e} \e \,f_1 \left(\D_i, \alpha_1 \right) f_3 \left(\D_i, \alpha_2 \right)~,
\end{align}
where 
\begin{multline}
   f_3 \left(\D_i, \alpha_2\right) \equiv \int_0^{\infty} dY \, Y^{\frac{\D_1-\D_3+\D_2}{2}}  \left|1-\alpha_2 Y \right|^{\D_3-2} \,  \text{sgn} \left(1 - \alpha_2 Y \right) \\ =\frac{1}{\pi}  \Gamma \left(\frac{2+\D_1+\D_2-\D_3}{2}\right) \Gamma \left(\frac{2-\D_1-\D_2-\D_3}{2}\right)
  \Gamma\left(\D_3 - 1\right) \\
 \times \begin{cases}
      -2\sin\frac{\pi}{2} \left(\D_1+\D_2\right) \cos \frac{\pi}{2}\D_3   &\alpha_2 = 1\\
         \sin \pi \D_3 &\alpha_2 = -1~.
    \end{cases}
\end{multline}
We can now put together the celestial amplitude corresponding to the $s$-channel exchange by summing over $\epsilon$ resulting in 
\begin{align}
\label{eq:Ascgr}
    \mAt_3^{sc,s} = -\frac{1}{2} \kappa_{2,2,-2}\kappa_{2,2,0} \, \mathcal{Z}\left(z_{ij}, \zb_{ij}\right) \, \mathcal{G}^{sc,s}\left(\D_i\right) \, \, \times\begin{cases}
    \mathcal{S}_{+} \quad(z_{12}\zb_{12}z_{13}\zb_{13}z_{23}\zb_{23}) \, > \, 0\\
    \mathcal{S}_{-}\quad(z_{12}\zb_{12}z_{13}\zb_{13}z_{23}\zb_{23}) \, < \, 0
    \end{cases}
\end{align}
where 
\begin{align}
\label{eq:Gscalar}
    \mathcal{G}^{sc,s}\left(\D_i\right) &= \frac{1}{\pi} \, \Gamma \left(\frac{2-\D_1-\D_2-\D_3}{2}\right) \Gamma \left(\frac{\D_1+\D_2-\D_3+2}{2}\right)   \\
    &\qquad \qquad \qquad \times \Gamma \left(\frac{\D_1-\D_2+\D_3}{2}\right) \Gamma \left(\frac{\D_2+\D_3-\D_1}{2}\right) 
    \frac{\Gamma\left(\D_3 - 1\right)}{\Gamma\left(\D_3\right)}~, \nonumber 
\end{align}
and $\mathcal{S}_{\pm}$ were defined in (\ref{eq:regionsgrav}). The invariance of $\mathcal{S}_{\pm}$ under cyclic permutations of the conformal dimensions $\Delta_i$ allows us to write the all-plus three-point expression in a compact manner by combining (\ref{eq:Atgr}) and (\ref{eq:Ascgr}) as follows: 
\begin{align}
    \an{\an{\mathcal{O}_{\D_1}^{++} \, \mathcal{O}_{\D_2}^{++} \, \mathcal{O}_{\D_3}^{++}}} &= \mAt_3^{gr} + \mAt_3^{sc,s} + \mAt_3^{sc,t} + \mAt_3^{sc,u}  \\
    &\nonumber= - \kappa_{2,2,-2}\kappa_{2,2,0} \, \mathcal{Z}\left(z_{ij}, \zb_{ij}\right) \, \left[\mathcal{G}^{gr} + \frac{1}{2}\left(\mathcal{G}^{sc,s} + \mathcal{G}^{sc,t} + \mathcal{G}^{sc,u}\right)\right] \\
    &\nonumber \qquad\qquad\times\begin{cases}
    \mathcal{S}_{+} \quad(z_{12}\zb_{12}z_{13}\zb_{13}z_{23}\zb_{23}) \, > \, 0\\
    \mathcal{S}_{-}\quad(z_{12}\zb_{12}z_{13}\zb_{13}z_{23}\zb_{23}) \, < \, 0~,
    \end{cases}
\end{align}
where $\mathcal{G}^{sc,t}(\Delta_i)$ and $\mathcal{G}^{sc,u}(\D_i)$ are obtained from $\mathcal{G}^{sc,s}(\D_i)$ given in (\ref{eq:Gscalar}) by interchanging $\D_1 \leftrightarrow \D_3$ and $\D_2 \leftrightarrow \D_3$ respectively. 

\subsubsection{$\an{\an{\mathcal{O}_{\D_1}^{++} \, \mathcal{O}_{\D_2}^{++} \, \mathcal{O}_{\D_3}^{--}}}$}
We now present the three-point function involving two positive and one negative helicity graviton derived from the momentum-space four-point amplitude given by (\ref{eq:gravppms}), which in the parameterization of (\ref{eq:masslessmompar22}) takes the form
\begin{align}
    \mA(1^{++},2^{++},3^{--},4^{\phi}) = 4^3 \kappa_{2,2,0} \kappa_{2,-2,-2} \frac{\zb_{12}^6 z_{23} z_{13}}{\zb_{23} \zb_{13}} \frac{ \e_1 \e_2 \om_1^3 \om_2^3}{4 \e_1 \e_2 \om_1 \om_2 z_{12} \zb_{12} - m^2}~. 
\end{align}
The computation of this Mellin transform is very similar to the previous cases and we omit all details. The final result takes the form
\begin{multline}
\label{eq:3ptfuncppm}
    \mAt_3\left(\left\lbrace\D_1, z_1, \zb_1\right\rbrace^{++},\left\lbrace\D_2, z_2, \zb_2\right\rbrace^{++},\left\lbrace\D_3, z_3, \zb_3\right\rbrace^{--}\right) \\
    = \kappa_{2,2,0} \, \kappa_{2,-2,-2} \, \tilde{\mathcal{Z}}\left(z_{ij},\zb_{ij}\right) \, \sum_{\e}  \e\, f_1 \left(\D_i, \alpha_1 \right) \, f_4 \left(\D_i, \alpha_2\right)~,
\end{multline}
where 
\begin{align}
\label{eq:conformalfactor2}
    \tilde{\mathcal{Z}} = 2^{-\beta-4}  m^{\beta+5} \, \frac{\zb_{12}^6 z_{23} z_{13}}{\zb_{23} \zb_{13}} \left|z_{12}\zb_{12}\right|^{\frac{\D_3-\D_1-\D_2-6}{2}} \left|z_{23}\zb_{23}\right|^{\frac{\D_1-\D_2-\D_3}{2}} \left|z_{13}\zb_{13}\right|^{\frac{\D_2-\D_1-\D_3}{2}}~,
\end{align}
is now the appropriate conformal factor, $f_1 \left(\D_i, \alpha_1\right)$ is the same integral as in (\ref{eq:Xint}) and 
\begin{equation} \label{eq:Yint2} \begin{aligned}
    f_4 \left(\D_i, \alpha_2\right) & \equiv \int_0^{\infty} dY \, Y^{\frac{\D_1-\D_3+\D_2}{2}+4}  \left|1-\alpha_2 Y \right|^{\D_3-2} \, \text{sgn} \left(1-\alpha_2 Y\right)  \\
   & = \frac{1}{\pi}  \Gamma \left(\frac{6+\D_1+\D_2-\D_3}{2}\right) \Gamma \left(\frac{-2-\D_1-\D_2-\D_3}{2}\right) \Gamma\left(\D_3 - 1\right) \\
   & \hspace{39mm} \times \begin{cases}
        -2\sin\frac{\pi}{2} \left(\D_1+\D_2\right) \cos \frac{\pi}{2}\D_3   &\alpha_2 = 1\\
         \sin \pi \D_3 &\alpha_2 = -1 \, .
         \end{cases}
\end{aligned} \end{equation}
Putting all of this together, the result is
\begin{align}
\an{\an{\mathcal{O}_{\D_1}^{++}\mathcal{O}_{\D_2}^{++}\mathcal{O}_{\D_3}^{--}}} = - \kappa_{2,-2,-2}\kappa_{2,2,0} \, \tilde{\mathcal{Z}}\left(z_{ij}, \zb_{ij}\right) \, \tilde{\mathcal{G}}\left(\D_i\right) \, \, \times\begin{cases}
    \mathcal{S}_{+} \quad(z_{12}\zb_{12}z_{13}\zb_{13}z_{23}\zb_{23}) \, > \, 0\\
    \mathcal{S}_{-}\quad(z_{12}\zb_{12}z_{13}\zb_{13}z_{23}\zb_{23}) \, < \, 0~,
    \end{cases}
    \label{eq:ppm3-pointfunc}
\end{align}
where 
\begin{align}
    \tilde{\mathcal{G}}\left(\D_i\right) &= \frac{1}{\pi} \, \Gamma \left(\frac{-2-\D_1-\D_2-\D_3}{2}\right) \Gamma \left(\frac{\D_1+\D_2-\D_3+6}{2}\right)  \nonumber \\
    &\qquad \qquad \qquad \times \Gamma \left(\frac{\D_1-\D_2+\D_3}{2}\right) \Gamma \left(\frac{\D_2+\D_3-\D_1}{2}\right) 
    \frac{\Gamma\left(\D_3 - 1\right)}{\Gamma\left(\D_3\right)}~,
\end{align}
and $\mathcal{S}_{\pm}$ were defined in (\ref{eq:regionsgrav}). 

\subsection{Four-point function}
The four-point function can be obtained from the five-point amplitude computed in Section \ref{sec:scamps} starting from the definition
\begin{equation}
\label{eq:5ptcadef}
     \mAt_4 \equiv \frac{m^2}{4} \int_0^{\infty} \frac{dy}{y^3} \, \int_{-\infty}^{\infty} \,dz_5\, d\zb_5 \int_0^{\infty} \prod_{i=1}^4 \frac{d\om_i}{\om_i} \, \om_i^{\D_i} \delta^{(4)} \left(\sum_{i=1}^5 p_i \right) \mA_5~.
\end{equation}
Due to the complexity of this amplitude, we will evaluate the corresponding celestial correlator only in Minkowski space. In this case, the regions on which the constraints imposed by the $\Theta$ functions are satisfied can be enumerated. As we will see below, a simplification occurs since $\zb_i$ are the complex conjugates of $z_i$ and we have $\zb_i z_i = \left| z_i \right|^2 >0$.\footnote{Note that in this section the symbols $z_i, \zb_i, z, \zb$ are all complex, and $\left| \right|$ denotes the modulus of the complex number. In the previous sections, they were real and we used $\left| \right|$ to denote the absolute value of these real numbers.} As before, we can first eliminate the coordinates of the scalar and one energy using
\begin{align}
    \label{eq:5ptdeltasolve}
    &\delta^{(4)} \left(\sum_{i=1}^5 p_i\right) = \frac{1}{|J|}\delta \left(\omega_4 - \omega_4^{\star}\right)\delta \left(z_5 - z_5^{\star}\right)\delta \left(\zb_5 - \zb_5^{\star}\right)\delta \left(y - y^{\star}\right)~,\\
    &\nonumber\omega_4^{\star} =  \frac{m^2 - 4 z_{12}\zb_{12} \e_1 \e_2 \om_1 \om_2 - 4 z_{13}\zb_{13}\e_1 \e_3 \om_1 \om_3 - 4 z_{23}\zb_{23} \e_2 \e_3 \om_2 \om_3}{4\e_4 \left(z_{14}\zb_{14}\e_1 \om_1+ z_{24}\zb_{24}\e_2 \om_2+z_{34}\zb_{34}\e_3 \om_3 \right)}~, \\
    &\nonumber J = -\frac{m^2 \e_5 \left(z_{14}\zb_{14}\e_1 \om_1+ z_{24}\zb_{24}\e_2 \om_2+z_{34}\zb_{34}\e_3 \om_3 \right) }{y^3}~,\\
    &\nonumber y^{\star} = \frac{-2m \e_5 \left(z_{14}\zb_{14} \e_1 \om_1 + z_{24}\zb_{24} \e_2 \om_2 + z_{34}\zb_{34} \e_3 \om_3\right)}{m^2+4\left(z_{14}\e_1 \om_1+z_{24}\e_2 \om_2+z_{34}\e_3 \om_3\right)\left(\zb_{14}\e_1 \om_1+\zb_{24}\e_2 \om_2+\zb_{34}\e_3 \om_3\right)}~.
\end{align}
In arriving at the above result, we have made use of the parametrization 
\begin{equation}
    p_i = \e_i \, \om_i \left( 1+z_i \zb_i, z_i + \zb_i , i \left(z_i - \zb_i \right), 1-z_i \zb_i \right)~,
\end{equation}
which is appropriate for Minkowski space. Performing the $y, z_5, \zb_5, \omega_4$ integrals, we get
\begin{multline}
\label{eq:5ptcamomsolv}
     \mAt_4 \equiv \int \prod_{i=1}^3 d\om_i \, \om_i^{\D_i-1} \frac{1}{|J|}\delta \left(\omega_4 - \omega_4^{\star}\right)\delta \left(z_5 - z_5^{\star}\right)\delta \left(\zb_5 - \zb_5^{\star}\right)\delta \left(y - y^{\star}\right) \\
    \times \left(\omega_4^{\star}\right)^{\D-4-1}\Theta \left(y^{\star}\right) \Theta \left(\omega_1^{\star}\right)\mA_5~.
\end{multline}
We do not sum over the directions as in the previous sections. We find it easier to solve the $\Theta$ function constraints and evaluate the amplitude for each case. Once we have these correlators, the summation can be trivially carried out. It is easy to see that the two $\Theta$ functions in (\ref{eq:5ptcamomsolv}) reduce to  
\begin{align}
    \label{eq:thetafuncs}
    \Theta \left(y^{\star}\right)\Theta \left(\om_4^{\star}\right) =& \Theta \left( -\e_5 \left(z_{14}\zb_{14} \e_1 \om_1 + z_{24}\zb_{24} \e_2 \om_2 + z_{34}\zb_{34} \e_3 \om_3\right)\right)\\
    \nonumber &\Theta \left(\e_4 \e_5 \left(-m^2 + 4 z_{12}\zb_{12} \e_1 \e_2 \om_1 \om_2 + 4 z_{13}\zb_{13}\e_1 \e_3 \om_1 \om_3 + 4 z_{23}\zb_{23} \e_2 \e_3 \om_2 \om_3 \right)\right).
\end{align}
The constraints imposed by these $\Theta$ functions on $\omega_1, \omega_2, \omega_3$ contours are most easily worked out by first changing variables to
\begin{align}
    \label{eq:varchange2}
    \om_1 = \frac{m}{2}\left|\frac{z_{34}}{z_{13}z_{14}}\right| \, \tilde{\om}_1, \,\,  \om_2 =\frac{m}{2}\left|\frac{z_{14}z_{34}}{z_{13}}\right|\frac{1}{|z_{24}|^2} \, \tilde{\om}_2,\,\,  \om_3 = \frac{m}{2}\left|\frac{z_{14}}{z_{13}z_{34}}\right| \, \tilde{\om}_3~.
\end{align}
The constraints can now be written purely in terms of the $\tom_i, \e_i$, the conformal cross ratio $z = \frac{z_{12}z_{34}}{z_{13}z_{24}}$ and its conjugate:
\begin{align}
    \label{eq:thetafuncssimple}
    \Theta \left(y^{\star}\right)\Theta \left(\om_4^{\star}\right) =& \Theta \left( -\e_5 \left(\e_1 \tom_1 +\e_2 \tom_2 + \e_3 \tom_3\right)\right)\\
    \nonumber &\Theta \left(\e_4 \e_5 \left(-1 + \e_1 \e_2 \left|z\right| \tom_1 \tom_2 + \e_1 \e_3 \tom_1 \tom_3 + \e_2 \e_3 \left|1-z\right| \tom_2 \tom_3 \right)\right)~.
\end{align}
Before enumerating all the regions, it helps to note that the $\Theta$ functions are invariant under three operations
\begin{enumerate}
    \item $\e_1 \tom_1 \leftrightarrow \e_2 \tom_2$
    \item $\e_1 \tom_1 \leftrightarrow \e_3 \tom_3 ~~ \text{and} ~ \left|z\right| \leftrightarrow \left|1-z\right|~$
    \item $\e_2 \tom_2 \leftrightarrow \e_3 \tom_3 ~~ \text{and} ~ \left|z\right| \leftrightarrow \left|1-z\right|~$.
\end{enumerate}
Moreover, we regard combinations of $\e_i$ differing by an overall sign to be equivalent and set $\e_5 = -1$ in everything that follows. These imply that we can obtain all possible regions from the ones shown in Table \ref{tab:regions}. The combinations $\e_1 = \e_2 = \e_3 = -\e_4 = 1$ and $\e_1 = \e_2 = \e_4 = -\e_3 = -1$ are not listed in this table as they lead to empty regions.\\ 
\begin{table}[htb!]
\centering
     \begin{tabular}{|l|l|l|l|l|L|}
      \hline
        Region & $\e_1$ & $\e_2$ & $\e_3$ & $\e_4$ & \text{Constraints} \\
        \hline
        $\mathcal{R}_1$ & 1 & 1 & 1 & 1 & \tom_1 >0, \quad \tom_2 <\frac{1}{\left|z\right|^2 \tom_1}, \quad 0<\tom_3 < \frac{1-\tom_1 \tom_2 \left|z\right|^2}{\tom_1+\tom_2 \left|1-z\right|^2}\\
        \hline
        $\mathcal{R}_2$ & 1 & 1 & -1 & 1 & \tom_1 >0, \quad \tom_2 <\frac{1}{\left|z\right|^2 \tom_1}, \quad 0<\tom_3 < \tom_1 + \tom_2\\
        \hline
        $\mathcal{R}_3$ & 1 & 1 & -1 & 1 &\tom_1 >0, \quad \tom_2 >\frac{1}{\left|z\right|^2 \tom_1}, \quad \frac{\tom_1 \tom_2 \left|z\right|^2-1}{\tom_1+\tom_2 \left|1-z\right|^2} < \tom_3 < \tom_1 + \tom_2\\
        \hline
        $\mathcal{R}_4$ & 1 & -1 & -1 & 1 & \tom_1>0, \quad 0<\tom_2 < \tom_1, \quad 0<\tom_3<\tom_1 - \tom_2\\
        \hline
        $\mathcal{R}_5$ & 1 & 1 & 1 & -1 &\tom_1 > 0, \quad 0 < \tom_2 < \frac{1}{\left|z\right|^2\tom_1}, \quad \tom_3 > \frac{1-\left|z\right|^2 \tom_1 \tom_2}{\tom_1 + \tom_2 \left|1-z\right|^2}\\
        \hline
        $\mathcal{R}_6$ & 1 & 1 & 1 & -1 &\tom_1 > 0, \quad \tom_2 > \frac{1}{\left|z\right|^2\tom_1}, \quad \tom_3 >0\\
        \hline
        $\mathcal{R}_7$ & 1 & 1 & -1 & -1 & \tom_1 > 0, \quad \tom_2 > \frac{1}{\left|z\right|^2\tom_1}, \quad \tom_3 < \frac{\left|z\right|^2 \tom_1 \tom_2 -1}{\tom_1 + \left|1-z\right|^2 \tom_2}\\
        \hline
\end{tabular}
\label{tab:regions}
     \caption{A list of regions carved out by the $\Theta$ function constraints for different values of $\e_i$. The regions for any combinations of $\e_i$ not listed here are either empty or are related to the listed ones via symmetries. In all of these cases we have set $\e_5 = -1$.}
\end{table}

The variable change (\ref{eq:varchange2}) also vastly simplifies (\ref{eq:5ptdeltasolve}) to 
\begin{align}
    &\omega_4^{\star} =-\frac{m \e_4}{2} \left|\frac{z_{13}}{z_{14}z_{34}}\right|\frac{\left(1-|z|^2 \e_1 \e_2 \tom_1 \tom_2 - \e_1 \e_3 \tom_1 \tom_3 - \left|1-z\right|^2 \e_2 \e_3 \tom_2 \tom_3\right)}{\left(\e_1 \tom_1 + \e_2 \tom_2 + \e_3 \tom_3 \right)}~,\\
    &\nonumber J = -\frac{m^3 \e_5}{2y^3}\left|\frac{z_{14}z_{34}}{z_{13}}\right|\left(\e_1 \tom_1 + \e_2 \tom_2 + \e_3 \tom_3 \right)~.
\end{align}
and the integral for the celestial correlator 
\begin{multline}
  \mAt_4 = \mathcal{Z}
    \int_{\mathcal{R}_i}  d\omega_1\, d\omega_2\, d\omega_3 \,\omega_1^{\D_1-1}\omega_2^{\D_2-1} \omega_3^{\D_3-1} \left(\e_1 \tom_1 + \e_2 \tom_2 + \e_3 \tom_3 \right)^{-\D_4}\\\left(1-|z|^2 \e_1 \e_2 \tom_1 \tom_2 - \e_1 \e_3 \tom_1 \tom_3 - \left|1-z\right|^2 \e_2 \e_3 \tom_2 \tom_3\right)^{\D_4-1} \mA_5^{\star}~.
\end{multline}
$\mA_5^{\star}$ is meant to represent the five-point amplitude after imposing momentum conservation and the variable change in (\ref{eq:varchange2}) and 
\begin{multline}
    \mathcal{Z} = \left|z_{34}\right|^{\frac{1}{2}\left(\D_1+\D_2-\D_3-\D_4-1\right)} \, \left|z_{13}\right|^{\frac{1}{2}\left(-\D_1-\D_2-\D_3+\D_4+1\right)} \\
    \left|z_{14}\right|^{\frac{1}{2}\left(-\D_1+\D_2+\D_3-\D_4+1\right)} \left|z_{24}\right|^{1-\D_2} \left(\frac{m}{2}\right)^{-5+\sum_{i=1}^4 \D_i}~,
\end{multline}
is the standard conformal factor.\\

It is prudent to first identify a few classes of integrals which arise in the computation of the correlator above. To this end, it is necessary to rewrite the five-point amplitude $(\ref{eq:5ptgravscalar})$ using the variables $(\ref{eq:varchange2})$ as in Appendix \ref{app:terms}. We can check that all the terms $\mA_5^{gr, 1,2 ,(i)}$ (defined in Appendix \ref{app:terms}) lead to the following class of integrals
\begin{multline}
    \label{eq:class1}
    I^{(1)}_{\mathcal{R}_i} \left(\alpha, \beta, \gamma, \delta, \rho, a,b,c\right) = \int_{\mathcal{R}_i} d\tom_1 \, d\tom_2 \, d\tom_3 \tom_1^{\alpha}\, \tom_2^{\beta} \, \tom_3^{\gamma} \left(1- \shat\right)^{\delta} \shat^a \left(\e_1 \tom_1 + \e_2 \tom_2 + \e_3 \tom_3 \right)^{\rho}\\
    \left(\shat \e_3 \tom_3+ \e_1 \tom_1 + \e_2 \tom_2\right)^b \left(\shat\left(\e_1 \tom_1 + \e_2 \tom_2\right) + \e_3 \tom_3\right)^c~,
\end{multline}
\begin{multline}
    \label{eq:class2}
    I^{(2)}_{\mathcal{R}_i} \left(\alpha, \beta, \gamma, \delta, \rho, a,b,c,d,e\right) \\
    = \int_{\mathcal{R}_i} d\tom_1 \, d\tom_2 \, d\tom_3 \tom_1^{\alpha}\, \tom_2^{\beta} \, \tom_3^{\gamma} \left(1- \shat\right)^{\delta} \left(\e_1 \tom_1 + \e_2 \tom_2 + \e_3 \tom_3 \right)^{\rho}  \left(\e_2 \tom_2 + \e_3 \tom_3 \right)^d\\
    \times\left(\shat\left(\e_2 \tom_2 + \e_3 \tom_3\right) + \e_1 \tom_1\right)^a \left(\shat \e_3 \tom_3 + \e_1 \tom_1 + \e_2 \tom_2\right)^b \\
    \times\left(1+\left|z\right|^2 \tom_2^2+\left(z+\zb\right)\e_2 \e_3 \tom_2\tom_3 + \tom_3^2\right)^c 
    \left(\e_2 \tom_2 \shat + \e_1 \tom_1 + \e_3 \tom_3\right)^e~,
\end{multline}
 where $\shat = |z| \e_1 \e_2 \tom_1 \tom_2 + \e_1 \e_3 \tom_1 \tom_3 + \left|1-z\right| \e_2 \e_3 \tom_2 \tom_3$. The exponents represented by Greek letters can take complex values while the Latin ones only take on integer values. All of the terms in $\mAt_4$ can be expressed in terms of these integrals. This decomposition can be performed easily using Mathematica. \\

The integrals in (\ref{eq:class1}, \ref{eq:class2}) must first be regulated with appropriate $i \epsilon$ prescriptions. Here, we must distinguish between the factors with Greek and Latin exponents. The former are positive in all the regions  $\mathcal{R}_i$ as they are just the arguments of the $\Theta$ functions. Consequently they do not require $i \epsilon$ prescriptions. However, the factors with integer exponents introduce new singularities which sometimes intersect the regions $\mathcal{R}_i$. The necessary $i \epsilon$ prescription  follows from the corresponding Feynman $i\epsilon$ for the propagators in the tree amplitude. We will now tacitly assume such a prescription and outline how we can derive a Mellin-Barnes representation for these integrals. Applying the Mellin-Barnes decomposition formula\footnote{Here and henceforth, we will suppress the contour of integration. It is understood that these contours run upward along the imaginary axis such that the poles separate  into left and right ones. For more details, please refer to \cite{Dubovyk:2022obc}.}
\begin{multline}
    \label{eq:MBdecomp}
    \frac{1}{(A_1+ \dots + A_n)^{\lambda}} = \frac{1}{(2 \pi i)^{n-1} \Gamma(\lambda)} \int \prod_{i=1}^{n-1} \, dx_i \, A_i^{x_i} A_n^{-\lambda-x_1-\dots-x_{n-1}} \\
\times \prod_{i=1}^{n-1} \Gamma(-x_i) \Gamma(\lambda+x_1+\dots+x_{n-1})~,
\end{multline}
to each factor with an integer exponent yields an integral of the form
\begin{align}
\label{eq:hardintMBrep}
    \mathcal{I}^{(j)}_{\mathcal{R}_i} = \int \prod_{k=1}^r dx_k\, G \left(x_1, \dots, x_r,a,b,c,d,e, z, \zb\right) \mathcal{J}_{\mathcal{R}_i}~.
\end{align}
This generic form is motivated by the observation that each of the factors $A_i$ in the Mellin-Barnes decomposition is a product of powers of $\tom_1, \tom_2, \tom_3, z$ and $\zb$.  This modifies the exponents of $\tom_1, \tom_2, \tom_3$ from $\alpha, \beta, \gamma$ in (\ref{eq:class1}, \ref{eq:class2}) to $A, B$ and $C$. The function $G \left(x_1, \dots, x_r, a,b,c,d,e\right)$ is a universal (independent of the regions $\mathcal{R}_i$) factor containing all the $\Gamma$ functions and powers of $z, \zb$ arising from (\ref{eq:MBdecomp}). $r$ is an integer which represents the total number of Mellin variables that must be introduced to achieve this form. We have also defined
\begin{equation}
\label{eq:hardint}
    \mathcal{J}_{\mathcal{R}_i} = \int_{\mathcal{R}_i} d\tom_1 \, d\tom_2 \, d\tom_3 \, \tom_1^{A} \, \tom_2^{B}\, \tom_3^{C} \left(1-\shat\right)^{\delta} \left(\e_1 \tom_1 + \e_2 \tom_2 + \e_3 \tom_3 \right)^{\rho}~.
\end{equation}
We can now complete the reduction to Mellin-Barnes form more efficiently by performing some of the integrals and identifying them with well known hypergeometric functions. The details of this somewhat lengthy procedure are relegated to Appendix \ref{sec:MBreduction}. Here, we merely present the final result:
\begin{equation}
    \label{eq:MBreducedform}
     \mathcal{J}_{\mathcal{R}_4} = \int \prod_{i=1}^4 \, dx_{r+i} \, \mathcal{G} \left(x_{r+1}, \dots, x_{r+4}\right)\left(z-1\right)^{x_{r+2}}\left(\zb-1\right)^{x_{r+3}}\left(1-\left|z\right|^2\right)^{x_{r+4}},
\end{equation}
where $\mathcal{G}$ is defined in \eqref{thefunctionG}.

\section{Graviton OPEs}
\label{sec:OPE}
We can obtain the graviton OPEs (for positive helicity gravitons) from the three-point correlator (\ref{eq:Atgr}, \ref{eq:Ascgr}). Since the OPE is local and is insensitive to insertions far away from the position of the relevant operators, we expect it to agree with the OPE derived for correlators without a massive scalar. Consider the limit as $z_{23} \to 0$ in which we have for the conformal factor 
\begin{equation}
    \mathcal{Z} \xrightarrow{2 \parallel 3} 2^{2-2\D_1} m^{2\D_1+2} \frac{\zb_{13}^2}{z_{13}^2}\frac{\zb_{23}}{z_{23}}\left|z_{13}\zb_{13}\right|^{-\D_1} \left(-i \pi\right) \delta\left(\D_2+\D_3-\D_1\right)~.
\end{equation}
In arriving at this result, we have made use of the identity~\cite{Donnay:2020guq}
\begin{equation}
    \lim_{\nu \to 0} \nu^{z-\D} \, \Gamma\left(\D-z\right) = -2\pi i \delta \left(\D-z\right)~.
\end{equation}
The terms in the three-point correlator now become
\begin{multline}
    \label{eq:OPElimitAgr}
\mAt_3^{gr} \xrightarrow{2 \parallel 3} -2i \kappa_{2,2,-2}\frac{\zb_{23}}{z_{23}}\frac{1}{\pi} \Gamma\left(1-\D_2-\D_3\right) \Gamma\left(\D_2\right) \Gamma\left(\D_3\right)\\
\times \left(\sin\pi \D_2 + \sin \pi \D_3 + \sin \pi\left(\D_2+\D_3\right)\right)\mAt_2 \\
=-2\kappa_{2,2,-2}\frac{\zb_{23}}{z_{23}}\Big(B(\D_2, \D_3) + B(1-\D_2-\D_3,\D_3)+B(1-\D_2-\D_3,\D_2)\Big)\mAt_2~,
\end{multline}

\begin{multline}
    \label{eq:OPElimitAscs}
\mAt_3^{sc,s} \xrightarrow{2 \parallel 3} -i \kappa_{2,2,-2}\frac{\zb_{23}}{z_{23}}\frac{1}{\pi} \Gamma\left(1-\D_2-\D_3\right) \Gamma\left(\D_2+1\right) \Gamma\left(\D_3-1\right)\\
\times \left(-\sin\pi \D_2 -\sin \pi \D_3 - \sin \pi\left(\D_2+\D_3\right)\right)\mAt_2\\
=-\kappa_{2,2,-2}\frac{\zb_{23}}{z_{23}}\Big(B(\D_2+1, \D_3-1) -B(1-\D_2-\D_3,\D_3-1)+B(1-\D_2-\D_3,\D_2+1)\Big)\mAt_2~,
\end{multline}

\begin{multline}
    \label{eq:OPElimitAscu}
\mAt_3^{sc,u} \xrightarrow{2 \parallel 3} -i \kappa_{2,2,-2}\frac{\zb_{23}}{z_{23}}\frac{1}{\pi} \Gamma\left(1-\D_2-\D_3\right) \Gamma\left(\D_3+1\right) \Gamma\left(\D_2-1\right)\\
\times \left(-\sin\pi \D_2 -\sin \pi \D_3 - \sin \pi\left(\D_2+\D_3\right)\right)\mAt_2\\
=-\kappa_{2,2,-2}\frac{\zb_{23}}{z_{23}}\Big(B(\D_2-1, \D_3+1) -B(1-\D_2-\D_3,\D_2-1)+B(1-\D_2-\D_3,\D_3+1)\Big)\mAt_2~,
\end{multline}
while $\mAt_3^{sc,t}$ is subleading in this limit and hence we drop it. Here $B(x,y)$ is the Euler beta function. Now, if we write the OPE as
\begin{equation}
\label{eq:OPE}
    \mo_{\D_2}^{++}\left(z_2, \zb_2\right) \mo_{\D_3}^{++}\left(z_3, \zb_3\right) \sim \frac{\zb_{23}}{z_{23}} C_{\D_2,\D_3}^{\D_2+\D_3} \, \mo_{\D_2+\D_3}^{++}\left(z_3, \zb_3 \right) + \dots~,
\end{equation}
we can read off the OPE coefficient to be
\begin{align}
    C_{\D_2,\D_3}^{\D} = -\kappa_{2,2,-2}\Big(&2B(\D_2, \D_3) + 2 B(1-\D_2-\D_3,\D_3) + 2B(1-\D_2-\D_3,\D_2) \nonumber \\
    &+B(\D_2+1, \D_3-1) - B(1-\D_2-\D_3,\D_3-1) - B(1-\D_2-\D_3,\D_2+1)\nonumber\\
    &+B(\D_2-1, \D_3+1) -B(1-\D_2-\D_3,\D_2-1) - B(1-\D_2-\D_3,\D_3+1)\Big)\nonumber\\
    =-\kappa_{2,2,-2}\Big(& B(\D_2-1, \D_3-1) -B(\D_2-1,3-\D_2-\D_3)-B(\D_3-1,3-\D_2-\D_3) \Big)~, 
    \label{eq:OPEcoeffcient}
\end{align}
where we have used the following identities:
\begin{itemize}
    \item $B(\D_2-1, \D_3+1) +B(\D_2+1,\D_3-1)+2B(\D_2,\D_3)-B(\D_2-1,\D_3-1) = 0$
    \item $B(1-\D_2-\D_3,\D_2-1) +B(1-\D_2-\D_3,\D_2+1)-2B(1-\D_2-\D_3,\D_2) \\~~~~~~~~~~~~~~~~~~~~~~~~~~~~~~~~~~~~~~~~~~~~~~~~~~~~~~~~~~~~~~~~-B(\D_2-1, 3-\D_2-\D_3) = 0$
    \item $B(1-\D_2-\D_3,\D_3-1) +B(1-\D_2-\D_3,\D_3+1)-2B(1-\D_2-\D_3,\D_3)\\~~~~~~~~~~~~~~~~~~~~~~~~~~~~~~~~~~~~~~~~~~~~~~~~~~~~~~~~~~~~~~~~-B(\D_3-1, 3-\D_2-\D_3) = 0~$,
\end{itemize}
to simplify the OPE coefficients in the final equality above. This OPE coefficient cannot be directly compared with those obtained in \cite{Stieberger:2018onx, Pate:2019lpp} as it does not correspond to OPEs of incoming or outgoing operators, but to specific linear combinations of them. We can use the OPEs of \cite{Stieberger:2018onx, Pate:2019lpp} to derive the appropriate quantity for comparison. To do this, we start by considering\footnote{We set $\e_2 =1$ since there are only two inequivalent choices for $\e_1, \e_2$ upto an overall sign.}\textsuperscript{,}\footnote{We suppress all the coordinate dependence of the operators.}
\begin{equation}
 \sum_{\e_3 = \pm 1} \mo_{\D_2}^{++,+1}\, \mo_{\D_3}^{++,\e_3} = \mo_{\D_2}^{++,+1} \,\mo_{\D_3}^{++,+1}+\mo_{\D_2}^{++,+1} \mo_{\D_3}^{++,-1} .
\end{equation}
Applying the usual OPE we get,
\begin{align}
  \sum_{\e_3 = \pm 1} \mo_{\D_2}^{++,+1}\, \mo_{\D_3}^{++,\e_3} \sim -\kappa_{2,2,-2}\frac{\zb_{23}}{z_{23}} \Big(&B(\D_2-1,\D_3-1)\, \mo_{\D_2+\D_3}^{++,+1}
\nonumber
\\&\quad-B(\D_2-1,3-\D_2-\D_3) \,\mo_{\D_2+\D_3}^{++,+1} \nonumber \\
&\qquad-B(\D_3-1,3-\D_2-\D_3) \mo_{\D_2+\D_3}^{++,-1}\Big)~.
\end{align}
If we now insert this into a three-point function and sum over the directions of the new insertion, we get
\begin{align}
   \sum_{\e_1,\e_3}\an{\an{\mo_{\D_1}^{++,\e_1}\, \mo_{\D_2}^{++,+1}\, \mo_{\D_3}^{++,\e_3} }}  \sim -\kappa_{2,2,-2} & \frac{\zb_{23}}{z_{23}} \Big(B(\D_2-1,\D_3-1) - B(\D_2-1,3-\D_2-\D_3) \nonumber \\
&- B(\D_3-1,3-\D_2-\D_3)\Big) \sum_{\e_1}\an{\an{\mo_{\D_1}^{++,\e_1}\mo_{\D_2+\D_3}^{++}}}~,
\end{align}
matching the expression for the OPE coefficient obtained in (\ref{eq:OPEcoeffcient}). This shows that the OPEs in the presence of a massive scalar are consistent with those in its absence. 
\section{The conformally soft graviton theorem}
\label{sec:conformalsofttheorem}
The computations of the previous sections have shown that celestial amplitudes involving $n$ gravitons and a massive scalar behave like $n$-point graviton correlators when the conformal dimension of the scalar is taken to vanish. We will now examine whether this similarly persists for the conformal soft theorems \cite{Puhm:2019zbl}. In this paper, we will only examine the leading soft theorem leaving the others for future work. In momentum space, the leading soft theorem \cite{Weinberg} for an amplitude with $n$ positive helicity gravitons and a massive scalar is 
\begin{align}
    \label{eq:leadingsofttheorem}
    \mA_{n+1} \left(\varepsilon p_1^{++}, p_2^{++}, \dots, p_{n}^{++},p_{n+1}\right) \xrightarrow[]{\varepsilon \to 0} \, \frac{1}{\varepsilon} S\left[p_1^{++}\right]  \mA_n \left(p_2^{++}, \dots, p_{n}^{++},p_{n+1}\right)  + \mathcal{O}\left(\varepsilon^0\right)~,
\end{align}
where $S\left[p_1^{++}\right]$ is the relevant soft factor given by
\begin{align}
    \label{eq:leadingsoftfactor}
    S\left[p_1^{++}\right]  = \sum_{i=2}^{n+1} \frac{\e_{\mu \nu}\left(p_1\right) p_i^{\mu}p_{i}^{\nu}}{p_i \cdot p_1}~,
\end{align}
and $\e_{\mu \nu}\left(p_1\right)$ is the polarization tensor of the soft graviton, which takes the form\footnote{In general, this can be written with two arbitrary spinors. Here, we have made a specific choice. The soft factor is, of course, independent of this choice.} 
\begin{align}
    \e^{\mu \nu}\left(p_1\right) = \frac{\lan{2}  \gamma^{\mu}\rsq{1}}{\an{2 1}} \frac{\lan{2} \gamma^{\nu} \rsq{1}}{\an{2 1}}~,
\end{align}
for a positive helicity graviton. Note that there is a sum over all the other particles involved in the amplitude. For the four-point amplitude $\mA_4 \left(1^{++}, 2^{++}, 3^{++}, 4^{\phi} \right)$ in (\ref{eq:4ptgravscalar}), this becomes 
\begin{align}
    S\left[p_1^{++}\right] &= \frac{\sq{12}}{\an{12}} \frac{\sq{13}}{\an{13}} \frac{\sq{23}^2}{t-m^2} = \frac{\zb_{12} \zb_{13} z_{23}^2}{z_{12} z_{13}} \frac{\om_2 \om_3}{2m \omega_1} \frac{y}{y^2+z_{14}\zb_{14}}.
    \label{eq:softfactor1}
\end{align}
It can be checked that the momentum-space amplitude (\ref{eq:4ptgravscalar}) satisfies the soft theorems. It is worth noting that the leading soft behavior arises entirely from the term in the amplitude that corresponds to the exchange of the massive scalar. Indeed, without this, the amplitude would fail to satisfy the soft graviton theorem as already stated in Section \ref{sec:scamps}. In order to examine the celestial avatar of this statement, we must compute
\begin{multline}
    \label{eq:celestialsoft1}
    \lim_{\D_1 \to 1} \left(\D_1 - 1\right) \tilde{\mA}_{n+1}^{\e} = \frac{m^2}{4}\underset{\omega_1 = 0}{\oint} d\omega_1  \int_0^{\infty} \prod_{i=2}^n \frac{d\omega_i}{\omega_i} \omega_i^{\D_i} \int_0^{\infty} \frac{dy}{y^3}\\
    \times \int_{-\infty}^{\infty} dz_{n+1} \, d\zb_{n+1} \, G_{\D_{n+1}} \left(z_{n+1},\zb_{n+1},w,\bar{w}\right) \mA_{n+1}
\end{multline}
where we have used the definition (\ref{eq:celampdef}) and the identity 
\begin{equation}
    \lim_{\varepsilon \to 0} \varepsilon \left|x\right|^{\varepsilon-1} = \delta \left(x\right),
\end{equation}
which holds for functions falling off at infinity at least as fast as $x^{-b}$ for some $b>0$. For the four-point amplitude under consideration, this turns into 
\begin{multline}
    \lim_{\Delta_1 \to 1} (\Delta_1-1) \,\tilde{\mA}_4 = \frac{m}{8} \frac{\zb_{12} \zb_{13} z_{23}^2}{z_{12} z_{13}}\int \frac{dy}{y^3} \, dz_4 \, d\zb_4  d\omega_2 \, d\omega_3 \, \omega_2^{\Delta_2} \, \omega_3^{\Delta_3} \\
    \times \left(\frac{y}{y^2+z_{45}\zb_{45}}\right)^{\Delta_4} 
   \left(\frac{y}{y^2+z_{14}\zb_{14}}\right) \mA_3~.
\end{multline}
We must now set $\D_4 =0$ in order to obtain the conformal soft theorem for the correlators considered in this paper. The presence of the soft factor prevents us from interpreting this quantity as a lower point celestial amplitude with $\D_4 = 0$. In fact, this can be interpreted as a new correlator with $\D_4 = 1$ and the scalar coordinates being $\left\lbrace z_1, \zb_1\right\rbrace$. Thus, we can write 
\begin{multline}
\label{eq:modifiedsoft}
     \lim_{\Delta_1 \to 1} (\Delta_1-1) \, \tilde{\mA}_4 \left(\left\lbrace \D_1, z_1, \zb_1\right\rbrace, \left\lbrace \D_2, z_2, \zb_2\right\rbrace, \left\lbrace \D_3, z_3, \zb_3\right\rbrace, \left\lbrace \D_4, w, \bar{w}\right\rbrace\right)\\
     = \frac{1}{2m} \frac{\zb_{12} \zb_{13} z_{23}^2}{z_{12} z_{13}} \tilde{\mA}_3 \left(\left\lbrace \D_2+1, z_2, \zb_2\right\rbrace, \left\lbrace \D_3+1, z_3, \zb_3\right\rbrace, \left\lbrace 1, z_1, \bar{z}_1\right\rbrace\right)~.
\end{multline}
This is also supported by the following direct computation
\begin{align}
\lim_{\Delta_1 \to 1} (\Delta_1-1) \mAt_3 &= \frac{m^2}{2} \frac{\zb_{12}^3 z_{23}^2}{z_{12}^3 z_{13}^2} \bigg|\frac{m^2}{4 z_{23} \zb_{23}}\bigg|^{\Delta_3} \bigg|\frac{m^2z_{13}\zb_{13}}{4z_{12}\zb_{12}z_{23}\zb_{23}}\bigg|^{\frac{\Delta_2-\Delta_3+1}{2}} \nonumber \\
    &~\qquad \qquad \times 
    \begin{cases}
        \frac{\pi}{\cos{\left(\frac{\pi}{2}(\Delta_2-\Delta_3)\right)}}~~;~~ \textrm{sgn}(z_{12}\zb_{12}z_{23}\zb_{23}) > 0 \\
        -\pi\tan{\left(\frac{\pi}{2}(\Delta_2-\Delta_3)\right)}~~;~~ \textrm{sgn}(z_{12}\zb_{12}z_{23}\zb_{23}) < 0~~.
        \label{eq:softlimitthenmellinint}
    \end{cases}
\end{align}
This suggests that the leading soft theorem relates the set of correlators with vanishing scalar conformal dimension to those with unit conformal dimension. It would be intriguing to investigate if this has an interpretation along the lines of \cite{Kapec:2022axw, Kapec:2022hih}. The connection between soft theorems and Ward identities of asymptotic symmetries in the presence of massive particles has been studied in \cite{Campiglia:2015kxa, Campiglia:2015qka} in general. It would also be interesting to see if any lessons could be extracted from this analysis.

\section{Outlook and discussion}
\label{sec:disc}
The correlators computed in this paper are well-defined and non-distributional, resembling ordinary CFT correlators. This computation opens up several lines of investigation. The first of these is understanding whether the remaining soft theorems are modified, along with their associated symmetries. Next is understanding the singularity structure of these correlators either using the Mellin Barnes representations (\ref{eq:MBreducedform}) and generalizing the analysis in \cite{Yuan:2018qva} or using the Euler integral representation (\ref{eq:class1}, \ref{eq:class2}) and methods in \cite{Mizera:2021icv, Caron-Huot:2021xqj}. A particularly interesting question is understanding how the massive exchange is manifest in the singularity structure. Another question which would benefit from further investigation is if an on-shell version of the correlators in \cite{Sleight:2023ojm} agree with those in this paper. This would provide a connection between the two different definitions. The final and perhaps most interesting and challenging question is whether we can identify a 2D CFT that reproduces the correlation functions in this paper.

\section*{Acknowledgements}
We are grateful to Shamik Banerjee, Eduardo Casali, Walker Melton, Sruthi Narayanan, Andrzej Pokraka, Andrea Puhm, Marcus Spradlin and Andy Strominger for helpful discussions. This work was supported in part by the US Department of Energy under contract {DE}-{SC}0010010 Task F, by Simons Investigator Award \#376208, and by a Bershadsky Distinguished Visiting Fellowship at Harvard (AV). AB was also supported by the Celestial Holography Initiative at the Perimeter Institute for Theoretical Physics and the Simons Collaboration on Celestial Holography. AYS was also supported by the STFC grant DRR00590.

\appendix 

\section{Derivation of the \texorpdfstring{$\mA(1^{++},2^{++},3^{--},4^{\phi})$}{\mA(1^{++},2^{++},3^{--},4^{\phi})} amplitude}
\label{app:++-phi}
This amplitude is not constructible via BCFW or three line shift recursions. The simplest way to derive this amplitude is by appealing to consistent factorization and locality along the lines of \cite{Arkani-Hamed:2017jhn}. The basic procedure is to compute the residues on all factorization channels by multiplying the appropriate three-point amplitudes and then guessing an expression that correctly reproduces all them. For the case at hand, we begin by computing the residue in the $u$-channel corresponding to a graviton exchange. This is
\begin{align}
\label{eq:4ptgravterm1-3}
    \underset{\sq{13}\to 0}{\text{Res}} \,\mA_4\left(1^{++},2^{++},3^{--},4^{\phi}\right) &=\frac{1}{\an{13}} \vcenter{\hbox{\includegraphics[scale=0.6]{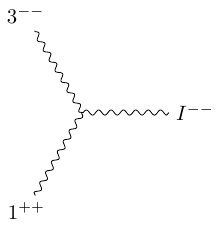} 
     \includegraphics[scale=0.6]{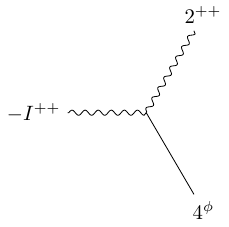}}}\\
     \nonumber &=\frac{1}{\an{13}}\kappa_{2,2,-2} \left(\frac{\an{3I}^3}{\an{I1}\an{13}}\right)^2 \, \kappa_{2,2,0}\,\sq{2I}^4\\
     &=\kappa_{2,2,0}\kappa_{2,2,-2} \frac{\sq{12}^6\an{23}^2\an{13}^2}{\an{13}t^2}~.
\end{align}
The expression above has a double pole in $t$ which would be unphysical in an amplitude. However, on the support of $u=0$, we have $s-m^2 = -t$. An ansatz for the amplitude is thus
\begin{equation}
    \mA_4 \left(1^{++},2^{++},3^{--},4^{\phi}\right)  
     =\kappa_{2,2,0}\kappa_{2,2,-2} \frac{\sq{12}^6\an{23}^2\an{13}^2}{t \, u \left(s-m^2\right)}~.
\end{equation}
It can be checked that this correctly reproduces the $t$- and $u$-channel residues and has the correct soft limits.

\section{Expressions for various terms in the five-point amplitude}
\label{app:terms}
This appendix contains expressions for all the terms in (\ref{eq:5ptgravscalar}) written in terms of the variables introduced in (\ref{eq:varchange2}). 
\begin{align}
    &\nonumber\mA_5^{gr,1,2, (1)} \equiv \frac{\sq{12}}{\an{12}}\frac{\sq{34}}{\an{34}}\frac{\sq{14}}{\an{14}}\frac{\sq{24}}{\an{24}} \frac{\left[3|p_5|4\right\rangle^2}{s_{24}} \frac{2}{s_{123}} = \frac{\zb}{z} \frac{\zb_{13}^2\zb_{24}^2}{z_{13}^2z_{24}^2} \e_2 \e_3 \frac{\tom_3}{\tom_2}  \frac{2\left(\e_1 \tom_1 + \e_2 \tom_2 \left(1-\zb\right) \right)^2}{\shat}\\
     &\mA_5^{gr,1,2, (2)} \equiv \frac{\sq{12}}{\an{12}}\frac{\sq{34}}{\an{34}}\frac{\sq{14}}{\an{14}}\frac{\sq{24}}{\an{24}} \frac{\left[3|p_5|4\right\rangle^2}{s_{14}} \frac{2}{s_{123}} = \frac{\zb}{z} \frac{\zb_{13}^2\zb_{24}^2}{z_{13}^2z_{24}^2} \e_1 \e_3 \frac{\tom_3}{\tom_1} \frac{2\left(\e_1 \tom_1 + \e_2 \tom_2 \left(1-\zb\right) \right)^2}{\shat}
\end{align}

\begin{align}
    &\nonumber\mA_5^{gr,1,2, (3)} \equiv -\frac{\sq{12}}{\an{12}}\frac{\sq{34}}{\an{34}}\frac{\sq{14}}{\an{14}}\frac{\sq{24}}{\an{24}} \frac{\left[3|p_5|4\right\rangle^2}{s_{24}}\frac{s_{34}}{s_{34}-m^2}\frac{1}{s_{123}} = \frac{\zb}{z} \frac{\zb_{13}^2\zb_{24}^2}{z_{13}^2z_{24}^2} \e_2 \frac{\tom_3^2}{\tom_2} \frac{\left(\e_1 \tom_1 + \e_2 \tom_2 \left(1-\zb\right) \right)^2}{\shat\e_3 \tom_3 +\e_1 \tom_1 +\e_2 \tom_2}\frac{\shat-1}{\shat}\\
    &\mA_5^{gr,1,2, (4)} \equiv -\frac{\sq{12}}{\an{12}}\frac{\sq{34}}{\an{34}}\frac{\sq{14}}{\an{14}}\frac{\sq{24}}{\an{24}} \frac{\left[3|p_5|4\right\rangle^2}{s_{14}}\frac{s_{34}}{s_{34}-m^2}\frac{1}{s_{123}} = \frac{\zb}{z} \frac{\zb_{13}^2\zb_{24}^2}{z_{13}^2z_{24}^2} \e_2 \frac{\tom_3^2}{\tom_1} \frac{\left(\e_1 \tom_1 + \e_2 \tom_2 \left(1-\zb\right) \right)^2}{\shat\e_3 \tom_3 +\e_1 \tom_1 +\e_2 \tom_2}\frac{\shat-1}{\shat}
\end{align}

\begin{align}    
   \nonumber \mA_5^{gr,1,2, (5)} &\equiv -\frac{\sq{12}}{\an{12}}\frac{\sq{34}}{\an{34}}\frac{\sq{14}}{\an{14}}\frac{\sq{24}}{\an{24}} \frac{\left[3|p_5|4\right\rangle^2}{s_{24}}\frac{s_{14}+s_{24}}{s_{14}+s_{24}-m^2}\frac{1}{s_{123}} \\
    \nonumber &=\frac{\zb}{z} \frac{\zb_{13}^2\zb_{24}^2}{z_{13}^2z_{24}^2} \e_2 \e_3 \frac{\tom_3}{\tom_2} \frac{\left(\e_1 \tom_1 + \e_2 \tom_2 \left(1-\zb\right) \right)^2\left(\e_1 \tom_1 + \e_2 \tom_2\right)}{\shat\left(\e_1 \tom_1 + \e_2 \tom_2\right) + \e_3 \tom_3}\frac{1-\shat}{\shat}\\
    \nonumber\mA_5^{gr,1,2, (6)} &\equiv -\frac{\sq{12}}{\an{12}}\frac{\sq{34}}{\an{34}}\frac{\sq{14}}{\an{14}}\frac{\sq{24}}{\an{24}} \frac{\left[3|p_5|4\right\rangle^2}{s_{14}}\frac{s_{14}+s_{24}}{s_{14}+s_{24}-m^2}\frac{1}{s_{123}} \\
    &=\frac{\zb}{z} \frac{\zb_{13}^2\zb_{24}^2}{z_{13}^2z_{24}^2} \e_2 \e_3 \frac{\tom_3}{\tom_1} \frac{\left(\e_1 \tom_1 + \e_2 \tom_2 \left(1-\zb\right) \right)^2\left(\e_1 \tom_1 + \e_2 \tom_2\right)}{\shat\left(\e_1 \tom_1 + \e_2 \tom_2\right) + \e_3 \tom_3}\frac{1-\shat}{\shat}
\end{align}

\begin{align}
    &\nonumber\mA_5^{gr,1,2, (7)} \equiv -\frac{\sq{12}}{\an{12}}\frac{\sq{34}}{\an{34}}\frac{\sq{14}}{\an{14}}\frac{\sq{24}}{\an{24}} \frac{\left[3|p_5|4\right\rangle^2}{s_{24}}\frac{1}{\shat-1}
    = \frac{\zb}{z} \frac{\zb_{13}^2\zb_{24}^2}{z_{13}^2z_{24}^2} \e_2 \e_3 \frac{\tom_3}{\tom_2} \frac{\left(\e_1 \tom_1 + \e_2 \tom_2 \left(1-\zb\right) \right)^2}{1-\shat}\\
    &\mA_5^{gr,1,2, (8)} \equiv -\frac{\sq{12}}{\an{12}}\frac{\sq{34}}{\an{34}}\frac{\sq{14}}{\an{14}}\frac{\sq{24}}{\an{24}} \frac{\left[3|p_5|4\right\rangle^2}{s_{14}}\frac{1}{\shat-1} 
    = \frac{\zb}{z} \frac{\zb_{13}^2\zb_{24}^2}{z_{13}^2z_{24}^2} \e_2 \e_3 \frac{\tom_3}{\tom_2} \frac{\left(\e_1 \tom_1 + \e_2 \tom_2 \left(1-\zb\right) \right)^2}{1-\shat}
\end{align}

\begin{align}
    \nonumber\mA_5^{sc, 1,2 (1)} &\equiv -\frac{\sq{23}\sq{34}\sq{24}\sq{14}\left[1|p_5|4 \right \rangle^2}{\an{23}\an{34}\an{24}\an{14}s_{14}}\frac{2}{s_{15}-s_{23}-m^2} \\
    &\nonumber=2\frac{\zb_{13}^2\zb_{24}^2}{z_{13}^2z_{24}^2}\frac{1-\zb}{1-z}\left(\zb \e_2 \tom_2+\e_3 \tom_3\right)^2\frac{\e_1 \tom_1 + \e_2 \tom_2 + \e_3 \tom_3}{\e_1 \tom_1 +\shat\left( \e_2 \tom_2 + \e_3 \tom_3\right)}\\
    \nonumber\mA_5^{sc, 1,2 (2)} &\equiv \frac{\sq{23}\sq{34}\sq{24}\sq{14}\left[1|p_5|4 \right \rangle^2}{\an{23}\an{34}\an{24}\an{14}s_{14}}\frac{2}{s_{15}-m^2}\\
    &=-2\frac{\zb_{13}^2\zb_{24}^2}{z_{13}^2z_{24}^2}\frac{1-\zb}{1-z}\left(\zb \e_2 \tom_2+\e_3 \tom_3\right)^2\e_1 \tom_1 \frac{\e_1 \tom_1+\e_2 \tom_2 + \e_3 \tom_3}{1+\left|z\right| \tom_2^2 + \left(z+\zb\right) \e_2 \e_3 \tom_2 \tom_3 + \tom_3^2}
\end{align}

\begin{align}
    \mA_5^{sc, 1,2 (3)} \nonumber&\equiv\frac{\sq{23}\sq{34}\sq{24}\sq{14}\left[1|p_5|4 \right \rangle^2}{\an{23}\an{34}\an{24}\an{14}s_{14}}\frac{1}{s_{15}-s_{23}-m^2}\frac{s_{34}}{s_{34}-m^2}\\
    &\nonumber=\frac{\zb_{13}^2\zb_{24}^2}{z_{13}^2z_{24}^2}\frac{1-\zb}{1-z} \frac{\e_3\tom_3\left(\zb \e_2 \tom_2+\e_3 \tom_3\right)^2\left(\e_1 \tom_1 + \e_2 \tom_2 + \e_3 \tom_3\right)\left(1-\shat\right)}{\left(\e_1 \tom_1 +\shat\left( \e_2 \tom_2 + \e_3 \tom_3\right)\right)\left(\shat \e_3 \tom_3+\e_1 \tom_1 + \e_2 \tom_2\right)} \\
    \nonumber\mA_5^{sc, 1,2 (4)} &\equiv -\frac{\sq{23}\sq{34}\sq{24}\sq{14}\left[1|p_5|4 \right \rangle^2}{\an{23}\an{34}\an{24}\an{14}s_{14}}\frac{1}{s_{15}-m^2}\frac{s_{34}}{s_{34}-m^2}\\
    &=-\frac{\zb_{13}^2\zb_{24}^2}{z_{13}^2z_{24}^2}\frac{1-\zb}{1-z}\frac{\left(\zb \e_2 \tom_2+\e_3 \tom_3\right)^2\e_1 \e_3 \tom_1\tom_3 \left(\e_1 \tom_1+\e_2 \tom_2 + \e_3 \tom_3\right)\left(1-\shat\right)}{\left(1+\left|z\right|^2 \tom_2^2 + \left(z+\zb\right) \e_2 \e_3 \tom_2 \tom_3 + \tom_3^2\right)\left(\shat \e_3 \tom_3+\e_1 \tom_1 + \e_2 \tom_2\right)}
\end{align}

\begin{align}
    \nonumber\mA_5^{sc, 1,2 (5)} &\equiv \frac{\sq{23}\sq{34}\sq{24}\sq{14}\left[1|p_5|4 \right \rangle^2}{\an{23}\an{34}\an{24}\an{14}s_{14}}\frac{1}{s_{15}-s_{23}-m^2}\frac{s_{123}+s_{14}}{s_{123}+s_{14}-m^2}\\
    \nonumber&=-\frac{\zb_{13}^2\zb_{24}^2}{z_{13}^2z_{24}^2}\frac{1-\zb}{1-z}\left(\zb \e_2 \tom_2+\e_3 \tom_3\right)^2\frac{\e_1 \tom_1 + \e_2 \tom_2 + \e_3 \tom_3}{\left(\e_2 \tom_2+\e_3 \tom_3\right)\left(\shat-1\right)}\\
    \nonumber\mA_5^{sc, 1,2 (6)} &\equiv -\frac{\sq{23}\sq{34}\sq{24}\sq{14}\left[1|p_5|4 \right \rangle^2}{\an{23}\an{34}\an{24}\an{14}s_{14}}\frac{1}{s_{15}-m^2}\frac{s_{123}+s_{14}}{s_{123}+s_{14}-m^2}\\
    &=\frac{\zb_{13}^2\zb_{24}^2}{z_{13}^2z_{24}^2}\frac{1-\zb}{1-z} \frac{\left(\zb \e_2 \tom_2+\e_3 \tom_3\right)^2\e_1 \tom_1\left(\e_1 \tom_1+\e_2 \tom_2 + \e_3 \tom_3\right)\left(\e_1 \tom_1 + \shat \left(\e_2 \tom_2+\e_3 \tom_3\right)\right)}{\left(1+\left|z\right|^2 \tom_2^2 + \left(z+\zb\right) \e_2 \e_3 \tom_2 \tom_3 + \tom_3^2\right)\left(\e_2 \tom_2+\e_3 \tom_3\right)\left(\shat-1\right)}
\end{align}

\begin{align}
    \nonumber\mA_5^{sc, 1,2 (7)} &\equiv \frac{\sq{23}\sq{34}\sq{24}\sq{14}\left[1|p_5|4 \right \rangle^2}{\an{23}\an{34}\an{24}\an{14}s_{14}}\frac{1}{s_{15}-s_{23}-m^2}\frac{s_{24}}{s_{24}-m^2}\\
    &\nonumber =\frac{\zb_{13}^2\zb_{24}^2}{z_{13}^2z_{24}^2}\frac{1-\zb}{1-z}\frac{\left(\e_1 \tom_1 + \e_2 \tom_2 + \e_3 \tom_3\right)\left(\zb \e_2 \tom_2+\e_3 \tom_3\right)^2\e_2\tom_2\left(1-\shat\right)}{\e_1 \tom_1 +\shat\left( \e_2 \tom_2 + \e_3 \tom_3\right)\left(\e_2 \tom_2 \shat+\e_1 \tom_1 + \e_3 \tom_3\right)}\\
   \nonumber \mA_5^{sc, 1,2 (8)} &\equiv -\frac{\sq{23}\sq{34}\sq{24}\sq{14}\left[1|p_5|4 \right \rangle^2}{\an{23}\an{34}\an{24}\an{14}s_{14}}\frac{1}{s_{15}-m^2}\frac{s_{24}}{s_{24}-m^2}\\
    &=-\frac{\zb_{13}^2\zb_{24}^2}{z_{13}^2z_{24}^2}\frac{1-\zb}{1-z}\frac{\left(\zb \e_2 \tom_2+\e_3 \tom_3\right)^2\e_1 \e_2 \tom_1 \tom_2\left(\e_1 \tom_1+\e_2 \tom_2 + \e_3 \tom_3\right)\left(1-\shat\right)}{1+\left|z\right|^2 \tom_2^2 + \left(z+\zb\right) \e_2 \e_3 \tom_2 \tom_3 + \tom_3^2\left(\e_2 \tom_2 \shat+\e_1 \tom_1 + \e_3 \tom_3\right)}~.
\end{align}

\section{Reduction of $\mathcal{J}_{\mathcal{R}_4}$ to a Mellin-Barnes integral}
\label{sec:MBreduction}
Start by changing variables to $\tom_1 = \tom_3 X, \tom_2 = \tom_3 Y$. This transforms (\ref{eq:hardint}) to 
\begin{multline}
    \label{eq:hardintstep1}
    \mathcal{J}_{\mathcal{R}_4} = \int_{0}^{\infty} dY \, \int_{1+Y}^{\infty} dX \, X^A\, Y^B \, \left(X-Y-1\right)^{\rho} \\
    \int_{0}^{\infty} d\tom_3 \, \tom_3^{A+B+C+2+\rho} \left[1+ \tom_3^2 \left(\left|z\right|^2 X Y + X - Y\left|1-z\right|^2\right)\right]^{\delta}~.
\end{multline}
We pause here to draw attention to the fact that the integral is well-defined as the integrand is strictly positive in the integration domain. The $\tom_3$ integral evaluates to 
\begin{multline}
    \label{eq:hardintom3}
     \int_{0}^{\infty} d\tom_3 \, \tom_3^{A+B+C+2+\rho} \left[1+ \tom_3^2 \left(\left|z\right|^2 X Y + X - Y\left|1-z\right|^2\right)\right]^{\delta} \\ = \mathcal{G}_1 \,  \left(\left|z\right|^2 X Y + X - Y\left|1-z\right|^2\right)^{-\frac{1}{2}\left(3+A+B+C+\rho\right)}~,
\end{multline}
with 
\begin{equation}
    \mathcal{G}_1 = \frac{1}{2\Gamma\left(-\delta\right)} \Gamma\left(-\frac{1}{2}\left(3+A+B+C+\rho+2\delta\right)\right)\Gamma\left(\frac{1}{2}\left(3+A+B+C+\rho\right)\right)~.
\end{equation}
After this, (\ref{eq:hardint}) reads,
\begin{multline}
    \mathcal{J}_{\mathcal{R}_4} = \mathcal{G}_1 \int_{0}^{\infty} dY \, \int_{1+Y}^{\infty} dX \, X^A\, Y^B \, \left(X-Y-1\right)^{\rho}\,  \left(\left|z\right|^2 X Y + X - Y\left|1-z\right|^2\right)^{-\frac{1}{2}\left(3+A+B+C+\rho\right)}.
\end{multline}
The evaluation of the integral over $X$ is expedited by the further change of variables, $t = \frac{-1+X-Y}{X-Y}$ as seen below:
\begin{multline}
\label{eq:hardintstep2}
    \mathcal{J}_{\mathcal{R}_4} = \mathcal{G}_1 \int_{0}^{\infty} dY \,Y^{B} \left(1+Y\right)^A  w_2^{-\frac{3+A+B+C+\rho}{2}} \\
   \times \int_{0}^1 dt \,t^{\rho}\, \left(1-t\right)^{-\frac{1+A+\rho-B-C}{2}}\left(1-\frac{Y}{Y+1}t\right)^A \left(1+\frac{w_1}{w_2}t\right)^{-\frac{3+A+B+C+\rho}{2}}.
\end{multline}
Here, $w_1 = Y(1-Y)\left|z\right|^2 - 2 Y\, Re(z), \, w_2 = 1+Y^2 \left|z\right|^2 + 2Y \, Re(z)$. We can recognize the integral over $t$ as a representation of the Appell $F_1$ function and write
\begin{multline}
    \int_{0}^1 dt \,t^{\rho}\, \left(1-t\right)^{-\frac{1+A+\rho-B-C}{2}}\left(1-\frac{Y}{Y+1}t\right)^A \left(1+\frac{w_1}{w_2}t\right)^{-\frac{3+A+B+C+\rho}{2}} \\
    = \mathcal{G}_2\, F_1 \left(\rho+1; -A, \frac{3+A+B+C+\rho}{2}; \frac{3-A+B+C+\rho}{2};\frac{Y}{Y+1}, -\frac{w_1}{w_2}\right)\\
    =\mathcal{G}_2 \left(\frac{w_1+w_2}{w_2}\right)^{-1-\rho} \hypf\left(-A, 1+\rho, \frac{1}{2}\left(3-A+B+C+\rho\right),\frac{w_1+Y(w_1+w_2)}{(w_1+w_2)(1+Y)}\right),
\end{multline}
with $\mathcal{G}_2 = \frac{\Gamma\left(\rho+1\right) \Gamma\left( \frac{1-A+B+C-\rho}{2}\right)}{\Gamma\left(\frac{3-A+B+C+\rho}{2}\right)} $ and we have used an identity of the Appell $F_1$ in arriving at the final equality. Plugging in the values of $w_1, w_2$, we see that the argument of the hypergeometric $\hypf$
\begin{align}
   0< \frac{w_1+Y(w_1+w_2)}{(w_1+w_2)(1+Y)} = \frac{Y \left|1-z\right|^2}{(1+y)(1+y \left|z\right|^2)}<1~.
\end{align}
This allows us to use the relevant Mellin-Barnes representation of the Gauss hypergeometric function,
\begin{multline}
    \label{eq:MBrep2F1}
    \hypf\left(a,b,c,z\right) = \frac{\Gamma(c)}{2\pi i \Gamma(a)\Gamma(b)\Gamma(c-a)\Gamma(c-b)} \\
    \times \int ds \, \Gamma(-s)\Gamma(c-a-b-s) \Gamma(a+s)\Gamma(b+s) \left(1-z\right)^{s}.
\end{multline}
In applying this formula, we introduce a new Mellin variable $x_{r+1}$ in addition to the $x_1, \dots, x_r$ introduced in (\ref{eq:hardintMBrep}). Thus, 
\begin{multline}
    \label{eq:hardintstep3}
    \mathcal{J}_{\mathcal{R}_4} = \int dx_{r+1} \,\mathcal{G}_3 \left(x_{r+1}\right)
    \int_{0}^{\infty} dY \,Y^{B} \left(1+Y\right)^{A-x_{r+1}}\left(1+y \left|z\right|^2\right)^{-1-\rho-x_{r+1}}\\
   \times \left(1+y \,z\right)^{-\frac{1}{2}\left(1+A+B+C-\rho-x_{r+1}\right)} \left(1+y \,\zb\right)^{-\frac{1}{2}\left(1+A+B+C-\rho-x_{r+1}\right)}~,
\end{multline}
where 
\begin{multline}
\mathcal{G}_3 \left(x_{r+1}\right) =  \Gamma(-x_{r+1}) \,\Gamma \left(-x_{r+1}+\frac{1}{2}(A+B+C-\rho)\right)\Gamma (-A+x_{r+1}) \\
\times \Gamma(\rho+1+x_{r+1})\frac{\Gamma\left(\frac{1}{2}(3+A+B+C+\rho+2\delta)\right)}{2\Gamma(-A)\Gamma(-\delta)}~.
\end{multline}
A final change of variables $Y = \frac{\eta}{1-\eta}$ turns the integral into
\begin{multline}
    \label{eq:hardintstep4}
    \mathcal{J}_{\mathcal{R}_4} = \int dx_{r+1} \,\mathcal{G}_3 \left(x_{r+1}\right)
    \int_{0}^{1} d\eta \,\eta^{B} \left(1-\eta\right)^{x_{r+1}+C}\left(1-\left(1-\left|z\right|^2\right)\eta\right)^{-1-\rho-x_{r+1}}\\
   \times \left(1 - \eta\left(1- z\right)\right)^{-\frac{1}{2}\left(1+A+B+C-\rho-x_{r+1}\right)} \left(1-\eta\left(1-\zb\right)\right)^{-\frac{1}{2}\left(1+A+B+C-\rho-x_{r+1}\right)}~,
\end{multline}
which can immediately be recognized as the integral representation of a type D Lauricella function,
\begin{equation}
    \label{eq:hardintstep5}
    \mathcal{J}_{\mathcal{R}_4} = \int dx_{r+1} \,\mathcal{G}_4 \left(x_{r+1}\right) F_D^{(3)} \left(a_1;b_1,b_2,b_3,c_1;1-z,1-\zb,1-\left|z\right|^2\right)~,
\end{equation}
where
\begin{align}
    &\nonumber\mathcal{G}_4\left(x_{r+1}\right) =\mathcal{G}_3\left(x_{r+1}\right) \frac{\Gamma\left(B+1\right)\Gamma\left(-x_{r+1}+C+1\right)}{\Gamma\left(-x_{r+1}+B+C+2\right)}~,\\
    &\nonumber a_1 = B+1, \quad b_1 = b_2 = \frac{1}{2}\left(1+A+B+C-\rho-x_{r+1}\right)~,\\
    & b_3 = 1+\rho-x_{r+1}, \quad c_1 =-x_{r+1}+B+C+2~.
\end{align}
To complete the reduction, we now use the Mellin-Barnes representation of the Lauricella function \cite{book:3578410}
\begin{multline}
    F_D^{(m)}\left(a_1;b_m, \dots , b_m; c_1,x_1, \dots , x_m\right) \\
    = \left(\frac{1}{2\pi i}\right)^m\int \prod_{i=1}^m dt_i \, \Gamma\left(-t_i\right)\left(-x_i\right)^{t_i}  \frac{\Gamma\left(c_1\right)\Gamma\left(a_1+\sum_{i=1}^m t_i\right)\prod_{i=1}^m\Gamma\left(b_i+t_i\right)}{\Gamma\left(a\right)\Gamma\left(c_1+\sum_{i=1}^m t_i\right)\prod_{i=1}^m \Gamma\left(b_i\right)}~, 
\end{multline}
to get the final Mellin-Barnes form
\begin{equation}
    \label{eq:hardintfinal}
     \mathcal{J}_{\mathcal{R}_4} = \int \prod_{i=1}^4 \, dx_{r+i} \, \mathcal{G} \left(x_{r+1}, \dots, x_{r+4}\right)\left(z-1\right)^{x_{r+2}}\left(\zb-1\right)^{x_{r+3}}\left(1-\left|z\right|^2\right)^{x_{r+4}}~,
\end{equation}
with
\begin{multline} \label{thefunctionG}
    \mathcal{G} \left(x_{r+1}, \dots, x_{r+4}\right) = \left[\prod_{i=1}^4 \Gamma\left(-x_{r+i}\right)\right]\frac{\Gamma\left(a_1+\sum_{i=2}^4x_{r+i}\right)}{\Gamma\left(c_1 +\sum_{i=2}^4 x_{r+i}\right)}\left[\prod_{i=2}^4 \frac{\Gamma\left(b_1+x_{r+i}\right)}{\Gamma\left(b_i\right)}\right]\Gamma\left(c_1-a_1\right)\mathcal{G}_3~.
\end{multline}

\bibliography{main}
\bibliographystyle{JHEP}
\end{document}